\documentclass[aps,preprint,nofootinbib,preprintnumbers,eqsecnum,superscriptaddress]{revtex4}

\usepackage{color}

\usepackage[normalem]{ulem}
\usepackage{enumerate}
\usepackage{amsfonts}
\usepackage{yfonts}

\usepackage{subfigure}
\usepackage{psfrag}

\usepackage{epsfig}
\usepackage[latin1]{inputenc}
\usepackage{float}
\usepackage{graphicx}
\usepackage{cancel}
\usepackage{mathrsfs}
\usepackage{amssymb}
\usepackage{amsfonts}
\usepackage{amsmath}
\usepackage{slashed}

\usepackage{braket}

\usepackage{graphicx}
\usepackage{bm}

\usepackage{bbold}

\usepackage{svg}

\usepackage[
      colorlinks=true,
      linkcolor=blue,
      urlcolor=blue,
      filecolor=black,
      citecolor=red,
      pdfstartview=FitV,
      pdftitle={},
        pdfauthor={Nabil Iqbal, Kieran Macfarlane},
        pdfsubject={},
        pdfkeywords={},
        pdfpagemode=UseNone,
		bookmarksopen=true,
		hyperfootnotes=false
	  ]{hyperref}

\newcommand \tr {\mbox{{\bf Tr}}}

\def\({\left(}
\def\){\right)}
\def\[{\left[}
\def\]{\right]}
\def\<{\langle}
\def\>{\rangle}

\newcommand{\bmat}{\begin{bmatrix}}
\newcommand{\emat}{\end{bmatrix}}

\def\Tr{\mathop{\rm Tr}}
\def\tr{\mathop{\rm tr}}

\newcommand\half{{\ensuremath{\frac{1}{2}}}}
\newcommand\p{\ensuremath{\partial}}

\newcommand{\be}{\begin{equation}}
\newcommand{\ee}{\end{equation}}
\newcommand{\bea}{\begin{eqnarray}}
\newcommand{\eea}{\end{eqnarray}}
\newcommand{\bwt}{\begin{widetext}}
\newcommand{\ewt}{\end{widetext}}

\newcommand{\bi}{\begin{itemize}}
\newcommand{\ei}{\end{itemize}}
\newcommand{\ben}{\begin{enumerate}}
\newcommand{\een}{\end{enumerate}}
\newcommand{\bca}{\begin{cases}}
\newcommand{\eca}{\end{cases}}
\newcommand{\bln}{\begin{align}}
\newcommand{\eln}{\end{align}}
\newcommand{\bst}{\begin{split}}
\newcommand{\est}{\end{split}}

\newcommand\al{{\alpha}}
\newcommand\ep{\epsilon}
\newcommand\sig{\sigma}

\newcommand\lam{\lambda}
\newcommand\Lam{\Lambda}
\newcommand\om{\omega}
\newcommand\Om{\Omega}

\newcommand\ga{{\ensuremath{{\gamma}}}}

\def\th{{\theta}}

\newcommand\ha{{\half}}

\def\le{\left}
\def\ri{\right}

\newcommand\sA{{\ensuremath{{\mathcal A}}}}

\newcommand\sM{{\ensuremath{{\mathcal M}}}}
\newcommand\sN{{\ensuremath{{\mathcal N}}}}

\newcommand\sP{{\mathcal P}}

\renewcommand{\Im}{\textrm{Im}\,}

\newcommand{\ka}{{\kappa}}

\newcommand{\NI}[1]{\textcolor{blue}{\textsf{[NI: #1]}}}

\begin{document}
\title {Higher-form symmetry breaking and holographic flavour}
\author{Nabil Iqbal}
\email{nabil.iqbal@durham.ac.uk}
\affiliation{Centre for Particle Theory, Department of Mathematical Sciences, Durham University,
South Road, Durham DH1 3LE, UK\\}
\author{Kieran Macfarlane}
\email{kieran.s.macfarlane@durham.ac.uk}
\affiliation{Centre for Particle Theory, Department of Mathematical Sciences, Durham University,
South Road, Durham DH1 3LE, UK\\}
\begin{abstract}
We study the realisation of higher-form symmetries in the holographic dual of gauge theories coupled to probe matter in the fundamental. We particularly focus on the dual of $U(N)$ gauge theory coupled to fundamental matter. We demonstrate the existence of a continuous 1-form symmetry associated with the conservation of magnetic flux and show that this symmetry is spontaneously broken in the IR when the flavour degrees of freedom are gapped. We numerically compute the spectral function of the 2-form current and demonstrate the existence of the associated Goldstone mode. We compare to expectations at weak-coupling. 

\end{abstract}
\vfill
\today

\maketitle
\tableofcontents
\pagebreak
\section{Introduction} \label{sec:intro}
In recent years, \textit{higher-form symmetries}, also called \textit{generalised global symmetries} \cite{Gaiotto:2014kfa}, have emerged as a powerful tool for understanding quantum field theory. Just as conventional global symmetries are associated with the conservation of densities of particles, higher-form symmetries are associated with the conservation of extended objects such as strings or flux tubes. They are on conceptually the same footing as ordinary global symmetries; they can spontaneously break (resulting in Goldstone modes \cite{Hofman:2018lfz,Lake:2018dqm}), they can have anomalies (see e.g. \cite{Gaiotto:2017yup} for an early example), and they can be used to build theories of hydrodynamics \cite{Grozdanov:2016tdf,Glorioso:2018kcp,Armas:2018atq}. 

In general, a $U(1)$ $p$-form symmetry is associated with a conserved $(p+1)$-form current $J$.
\begin{equation}
	d \star J = 0	
\end{equation}
The conventional case is $p = 0$; in this case we have a 1-form current that counts a density of particles. Here we will focus on the case $p =1$; in this case we have a conserved $2$-form current satisfying $\partial_\mu J^{\mu \nu} = 0$.  

In this work we will study some aspects of the realisation of higher-form symmetries in quantum field theories with holographic duals. We will focus on the study of gauge theory coupled to probe matter in the fundamental representation. As we will review below, in the case where the gauge group is $SU(N)$, this theory has no 1-form symmetries, though it does have a 0-form symmetry associated with baryon number; we will clarify some aspects of how the holographic representation of this baryon number intertwines with the (explicitly broken) putative center symmetry of the pure gauge theory. In the case where the gauge group is $U(N)$ however, the theory has an unbroken 1-form symmetry associated with the conservation of magnetic flux of the ``U(1) factor'' in the gauge group. We will study the realisation of this symmetry, identifying the charged line operators and studying the correlation function of its currents. As we will elaborate on below, we note that this is perhaps the simplest holographic model in which such a continuous 1-form symmetry can be spontaneously broken, motivating our study. 

In this leisurely introduction we begin by reviewing how higher-form symmetries are realised in various types of non-Abelian gauge theory with and without matter couplings. 

\subsection{Higher-form symmetries in non-Abelian gauge theory}
{\bf $SU(N)$ gauge theory:} Let us begin our study by reviewing the higher-form symmetry structure of $SU(N)$ gauge theory with only adjoint matter. If we have access to a Lagrangian description of the theory, the action is
\begin{equation}
S = \int d^4x \le(- \frac{1}{g_{YM}^2} \tr |F|^2 + \cdots \ri) \label{gaugeac} 
\end{equation}
where $F$ is the non-Abelian field strength, and the $\cdots$ refers to possible supersymmetrisations or other terms in the action. This action depends only on the Lie algebra of the group $su(N)$. As it turns out, though the Lie algebra specifies the action, it does not actually {\it fully} define the theory itself. This is because the full theory contains line operators, and the spectrum of line operators depends on the global form of the gauge group \cite{Aharony:2013hda}. Let us first consider the case where the global form of the gauge group is $SU(N)$; we then have the usual Wilson lines in the fundamental representation of the gauge group in the theory:
\begin{equation}
W(C) \equiv \Tr\mathrm{P} \exp\le(\oint_{C} A\ri)
\end{equation}

In a modern understanding, these Wilson lines are charged under 1-form symmetries. To be more precise, there is a $\mathbb{Z}_N$ valued surface operator $U_{q}(\sM_2)$ that is defined on closed 2-manifolds and is {\it topological} in that it is invariant under small modifications of the 2-manifold $\sM_2$. This surface operator has a non-trivial braiding algebra with the Wilson line, i.e. we have inside the path integral,
\begin{equation}
U(\sM_2) \; W(C)  = \exp\le(\frac{2 \pi i q}{N}\ri) W(C) \qquad q \in \{0,\ldots,N-1\}
\end{equation}
if $\sM_2$ wraps the curve $C$. This 1-form symmetry is a refinement of the usual ``center'' symmetry of non-Abelian gauge theory, under which adjoint matter fields $\Phi^a_b$ are invariant:
\begin{equation}
	\Phi^a_b \to \exp\left(\frac{2 \pi i q}{N}\right) \delta^a_c \; \Phi^c_d \; \exp\left(-\frac{2 \pi i q}{N}\right) \delta^d_b = \Phi^a_b
\end{equation}
(More explicitly, the insertion of a surface operator $U(\sM_2)$ induces a gauge transformation that is not single-valued as one winds around $\sM_2$; instead this gauge transformation returns to itself only up to an element of the center of the gauge group \cite{tHooft:1977nqb}. However if all fields transform in the adjoint, this operation is non-singular from the point of view of the gauge fields). 

Let us now consider what happens if we instead couple this theory to $N_f$ flavours of bosons and fermions charged in the fundamental under $SU(N)$, i.e. if the action is taken to be:
\begin{align}
S' = \int d^4x \; \left[- \frac{1}{ g_{YM}^2} \tr |F|^2 + \sum_{i = 1}^{N_f} \le(-|\p \phi_i - i A \phi_i|^2 - m_{\phi}^2 \; |\phi_i|^2 + \bar\psi_i \le(i \ga^{\mu} \le(\p_{\mu} - i A_{\mu} \ri)-m_{\psi}\ri)\psi_i\ri) + \cdots\right] \label{deformLag1} 
\end{align}
The 1-form symmetry above is now explicitly broken -- though the line operator $W(C)$ can still be defined, the operator $U_{q}$ is no longer topological and thus there is no longer a 1-form symmetry.\footnote{In familiar $0$-form language, a fundamental matter field transforms as \\
$\phi^a \to \exp\left(\frac{2 \pi i q}{N}\right) \delta^a_b \; \phi^b =  \exp\left(\frac{2 \pi i q}{N}\right) \phi^a \neq \phi^a$.}

However there is a new $0$-form symmetry: the baryon number current, which acts as a diagonal phase rotation on both $\phi_i$ and $\psi_i$. The associated conserved current is defined in the usual way as:
\begin{equation}
j^{\mu}_B = \sum_{i = 1}^{N_f} \le(\bar{\psi}_{i} \ga^{\mu} \psi_i + 2 \; \Im \phi_i^{\dagger} D^{\mu} \phi_i \ri) \label{baryonnum} 
\end{equation}
It is worth noting that the local gauge-invariant operators that are charged under this baryon number symmetry are fully antisymmetrized products of $N_c$ fundamental fields, and so will have charge $N_c$ in the appropriate units. 

{\bf Symmetry structure of $U(N)$ gauge theory:} 
Let us now change the theory under consideration by studying instead the $U(N)$ gauge theory with only adjoint matter. It is convenient to write the gauge group as
\begin{equation}
U(N) = \frac{U(1) \times SU(N)}{\mathbb{Z}_N}
\end{equation}
If we have access to a Lagrangian description of the $U(N)$ gauge theory, it is straightforward to see that the gauge field corresponding to the $U(1)$ factor separates off, and the action \eqref{gaugeac} can now be written as
\begin{equation}
S = \int d^4x \le(-\frac{1}{2 g_1^2} |f|^2 - \frac{1}{g_{YM}^2} \tr |F|^2 + \cdots \ri)  \label{freeEM} 
\end{equation}
where $f$ corresponds to the field strength of the new $U(1)$ gauge field $f = da$. As all matter is in the adjoint, nothing couples to the $U(1)$ gauge field, which has a free Maxwell action. There is thus a precisely marginal $U(1)$ gauge coupling which we have named $g_1$.

Unlike above, where we had only a single discrete $\mathbb{Z}_N$ 1-form symmetry, this theory has two continuous $U(1)$ 1-form symmetries corresponding to the simultaneous conservation of electric $U(1)_e$ and magnetic $U(1)_b$ flux. Their respective conserved currents $J^{\mu\nu}_{e,b}$ are:
\begin{equation}
J^{\mu\nu}_e = \frac{1}{g_1^2} f^{\mu\nu} \qquad J^{\mu\nu}_b = \ha \ep^{\mu\nu\rho\sig} f_{\rho\sig} \label{1formcurrs} 
\end{equation}
In the phase described by a free $U(1)$ gauge theory action both of these symmetries are spontaneously broken, and the usual 4d photon is the Goldstone mode of this breaking. The line operator that is charged under the electric 1-form symmetry is the usual $U(1)$ Wilson line\footnote{This can be easily understood by applying the usual Noether procedure to the action \eqref{freeEM}, where the symmetry in question is realised by a shift of the $U(1)$ gauge potential by a closed 1-form $\Lam$: $a \to a + \Lam$.}, and that charged under the magnetic 1-form symmetry is the t'Hooft line. The diagnosis of this symmetry breaking in terms of these line operators is discussed in \cite{Hofman:2018lfz,Lake:2018dqm}. 

Following along the lines of the discussion above, we now add the same flavour degrees of freedom to the $U(N)$ gauge theory. The $U(1)$ gauge field $a$ now couples minimally to both $\phi$ and $\psi$:
\begin{align}
S' = \int d^4x \; \Bigg[-& \frac{1}{2 g_1^2} |f|^2 - \frac{1}{ g_{YM}^2} \tr |F|^2 + \nonumber \\
& \sum_{i = 1}^{N_f} \le(-|\p \phi_i - i a \phi_i - i A \phi_i|^2 - m_{\phi}^2 \; |\phi_i|^2 + \bar\psi_i \le(i \ga^{\mu} \le(\p_{\mu} - i a_{\mu} - i A_{\mu} \ri) - m_{\psi}\ri)\psi_i\ri) + \cdots\Bigg] \label{deformLag2} 
\end{align} 
where $a$ and $A$ are the $U(1)$ and $SU(N)$ gauge potentials respectively. This changes the dynamics of the $U(1)$ 1-form symmetries described above. Importantly, the symmetry corresponding to conservation of electric flux is now generically {\it explicitly broken} by the presence of the electrically charged matter; the simplest way to see this is to note that the conserved current identified in \eqref{1formcurrs} is now no longer conserved; indeed the $U(1)$ Maxwell equations are simply:
\begin{equation}
\p_{\nu} J^{\mu\nu}_e = j_B^{\mu}[\phi, \psi], \label{efluxnoncons} 
\end{equation}
where $j_B^{\mu}[\phi, \psi]$ is precisely the baryon number current \eqref{baryonnum}. Physically, this simply captures the idea that electric field lines can now end on the electric charges that are carried by $\phi$ and $\psi$. 

The magnetic flux current $J^{\mu\nu}_b$ is still conserved, as is clear from its definition:
\begin{equation}
\p_{\nu} J^{\mu\nu}_b = 0
\end{equation}
Thus, this theory has a single $U(1)$ 1-form symmetry. 

However, the realisation of this symmetry now depends on the dynamics of the $\phi, \psi$ fields. Let us now consider how the energy scale of interest $E$ compares to the masses $m_{\phi}, m_{\psi}$:
\ben \item $E \ll m_{\phi, \psi}$: In this case the matter fields are gapped and can essentially be ignored. We are then in the same situation as when there were no flavour fields at all; $j^{\mu}[\phi, \psi]$ is effectively zero, and both the electric and magnetic flux currents are conserved. The associated symmetries are again both {\it spontaneously} broken, as described around \eqref{1formcurrs}. In particular, the relevant line operators should display a perimeter law in this phase. 

\item $E \gg m_{\phi, \psi}$: In this case we probe electric charge fluctuations in the vacuum, and the electric flux symmetry is explicitly broken. The magnetic flux symmetry is now realised differently; in particular, it is no longer spontaneously broken. Relatedly, in this regime the $U(1)$ gauge coupling $g_1$ runs logarithmically with the energy scale $E$. 
\een

{\bf Summary:} In this work, we will study the manifestation of the higher-form symmetry structures described above in a strongly coupled model, given by the holographic realisation of maximally supersymmetric $\sN = 4$ Super-Yang-Mills coupled to matter in the fundamental. We will primarily focus on the case of the $U(N)$ gauge theory where we have a continuous 1-form symmetry, but along the way we will clarify some aspects of the $SU(N)$ case as we proceed. 

 An outline of the rest of the paper follows. In Section \ref{sec:symms} we introduce the (well-known) bulk action and discuss its symmetry structure, also discussing some lower-dimensional examples to build some intuition for the extensive dualisations that follow. In Sections \ref{sec:SU} and \ref{sec:U} we discuss the bulk dynamics and appropriate charged operators in the duality frames that are appropriate for the $SU(N)$ and $U(N)$ gauge theories respectively. Finally in Section \ref{sec:flucspec} we numerically compute the spectral function for the 2-form current in the $U(N)$ theory and compare with expectations at weak coupling.

\section{Symmetries of holographic flavour} \label{sec:symms} 
In this section we describe the holographic dual of the system described above; in particular we study the maximal supersymmetrisation of the gauge theory, i.e. $\sN = 4$ SYM with holographic flavour added. In most discussions of holography it is implicitly assumed that the gauge group is $SU(N)$; for us however the precise distinction between the $U(N)$ and $SU(N)$ gauge theories will be of considerable importance. This issue has been clarified recently (see \cite{Hofman:2017vwr} for a perspective from higher-form symmetry) and we briefly review it here, taking special care with the issues that will be relevant for our construction. 
\subsection{Bulk holographic action}
The holographic dual of the above is Type IIB string theory on $\text{AdS}_5 \times S^5$, giving rise to kinetic terms for the NS-NS $2$-form $B_2$ and the R-R $2$-form $C_2$, as well as a Chern-Simons term. After compactifying on the $S^5$ we obtain an action which is an integral over all of $\text{AdS}_5$. To add fundamental matter, we wrap $N_f \ll N_c$ probe $D7$-branes around the $S^5$ \cite{Karch:2002sh}. See e.g. \cite{CasalderreySolana:2011us} for a review of holographic flavour. 

The final form of the dimensionally-reduced action on $\text{AdS}_5$ is
\begin{equation}
	\begin{split} 
		S_{\text{bulk}}
		& = S_{\text{kin}} + S_{\text{CS}} + N_f \; S_{\text{DBI}} \\
		& =  \frac{N_c^2}{8 \pi^2 R^3}\int{\left[- \frac{1}{2} H_3^2 - \frac{1}{2}\left(\frac{\lambda}{4 \pi N_c}G_3 \right)^2 + \kappa B_2 \wedge \left(\frac{\lambda}{4 \pi N_c}G_3 \right) - \frac{1}{2} \kappa^2 \mu f(z)\left(B_2 + \frac{2 \pi R^2}{\sqrt{\lambda}} F_2 \right)^2 \right]} \label{bulkac} 
	\end{split}
\end{equation}

Note our conventions for writing differential forms. For a $p$-form $\Omega_p$ in $n$-dimensions we can define a corresponding $n$-form by
\begin{equation}
	\Omega_p^2 \equiv \Omega_p \wedge \star \Omega_p
\end{equation}
Sometimes it is preferable to work in components, in which case we borrow from \cite{Polchinski:1998rr} and write
\begin{equation}
	|\Omega_p|^2 \equiv \frac{1}{p!} \; \Omega_{\mu_1 \mu_2 \ldots \mu_p} \Omega^{\mu_1 \mu_2 \ldots \mu_p}
\end{equation}
It is straightforward to translate between these descriptions using the identity
\begin{equation}
	\Omega_p^2 = |\Omega_p|^2 \sqrt{|g|} \; d^n x
\end{equation}
Further conventions and notation are explained in Appendix \ref{conventions-appendix}. 

The forms appearing in the action are the field strengths $H_3 = dB_2$, $G_3 = dC_2$ and $F_2 = dA_1$. Note that we have used the unusual name $G_3$ for the field strength of the R-R form to avoid confusion with the field strength of the $D7$-brane Maxwell field. Our zoo of higher-form fields is extensive -- and will become even more so as we dualise fields in the bulk -- so we have provided an index in Appendix \ref{symbol-index}. The constants in the action are given by
\begin{subequations}
	\begin{align}
		\kappa	& = \frac{4}{R}						\label{kappa-def}	\\
		\mu		& = \frac{N_f}{N_c}\frac{\lambda}{32 \pi^2}	\label{mu-def}
	\end{align}
\end{subequations}
$\mu$ denotes the relative dynamical importance of the flavour and colour degrees of freedom. The function $f$ is given by
\begin{equation}
	\label{f-def}
	f(z) = 
	\begin{cases} 
	1 - \left(z/z_c\right)^2	& z\leq z_c \\
	0				& z > z_c
	\end{cases}
\end{equation}

We will work with $\text{AdS}_5$ in Poincar\'e coordinates:
\begin{equation}
ds^2 = \frac{R^2}{z^2} \left(dx^{\mu} dx_{\mu} + dz^2\right)	\label{Poincare-coords}
\end{equation}
where $R$ is the AdS radius.

Some words about the probe limit are in order here. Usually one considers the limit where $\mu \to 0$, and thus where the backreaction of the flavour degrees of freedom on the colour degrees of freedom can be ignored. We will work with the simple quadratic action above, but we will allow $\mu$ to take on finite values. This corresponds to studying some subset of the interplay between flavour and colour degrees of freedom, in particular those associated with the realisation of the symmetries. As explained below, this results in novel effects associated with the Higgsing of the 2-form $B$ field by the DBI gauge field $A_1$; these effects qualitatively affect the physics but are invisible in the strict probe limit.  Strictly speaking, however, we are not studying all aspects of this interplay, because we neglect the gravitational backreaction of the flavour branes; thus the approximation we take should be considered as an illustrative one that is designed to highlight the physics of interest. 

Let us now note the gauge symmetry of the action above. We have two independent 1-form gauge transformations shifting $B_2$ and $C_2$ respectively; we also have a 0-form gauge transformation shifting the DBI worldvolume field. The full transformation of the fields is
\begin{subequations}
	\begin{align}
		\delta B_2 &= d \Xi^{(B)}_1				\\
		\delta C_2 &= d \Xi^{(C)}_1				\\
		\delta A_1 &= - \frac{\sqrt{\lambda}}{2 \pi R^2} \; \left(\Xi^{(B)}_1 + d\xi^{(A)}\right)
	\end{align} 
\end{subequations}
Note the simultaneous transformation of $A_1$ and $B_2$ under a shift by $\Xi^{(B)}_1$; this encodes the fact that string worldsheets can end on the $D$-brane, and will be of considerable importance to us. 


We turn now to the Chern-Simons term $B_2 \wedge G_3$; this well-known term \cite{Aharony:1998qu, Maldacena:2001ss, Witten:1998xy} is closely related to the physics of higher-form symmetry in holography \cite{Hofman:2017vwr} and will play a key role in our analysis. Obtaining the precise prefactor can be somewhat subtle; it can naively be thought of as arising from a dimensional reduction of a 10d Chern-Simons term $B_2 \wedge dc_4 \wedge G_3$ involving the R-R $4$-form $c_4$. Integrating over the $S^5$ we pick up a factor of the flux $dc_4 \sim N_c$, giving a term with qualitatively the correct form, as first noted in \cite{Witten:1998xy}. 

However, this is not quite consistent: in fact, a covariant action for the (self-dual) RR 4-form does not actually exist, and pursuing the above route results in an inconsistent normalisation for the Chern-Simons term, as noted in \cite{Belov:2004ht}. In this work we take a different approach. Consistency of the theory in the presence of magnetic charges actually requires the coefficient of this term to be quantized; we review this Dirac quantisation condition in Appendix \ref{Chern-Simons appendix} and identify the integer coefficient of the term with $N_c$, as we expect on symmetry grounds.

Finally, the last term arises from the dimensional reduction of the DBI action. We can embed $N_f$ probe $D7$-branes into the target space by means of the DBI action:
\begin{equation}
	N_f \; S_{\text{DBI}} = - N_f\; \tau_7 \int{d^8\xi \; \sqrt{-\det\left(g_{\alpha \beta} + B_{\alpha \beta} + 2 \pi l_s^2 \; F_{\alpha \beta} \right)}}
\end{equation}
where $\xi^\alpha$ are the brane worldvolume coordinates, $g_{\alpha \beta}$ are the components of the induced worldvolume metric on the $D7$-brane, $B_{\alpha \beta}$ are the components of $B_2$ and $F_{\alpha \beta}$ are the components of $F_2 = dA_1$, the Maxwell field strength living on the brane worldvolume. $\tau_p$ is the effective $Dp$-brane tension after absorbing the effect of the dilaton $e^\Phi = g_s$ and is given by $\tau_7 = \frac{1}{g_s}\frac{1}{(2\pi)^7 l_s^8}$. 

The $S^5$ factor in the metric may be written as
\begin{equation}
d\Omega_5^2 = d\theta^2 + \cos^2{\theta} \; d\psi^2 + \sin^2{\theta} \; d\Omega_3^2,
\end{equation} 
where we have chosen coordinates that make manifest an $S^3 \subset S^5$. As usual for a D3/D7 embedding, the desired brane configuration fills all of $\text{AdS}$ and wraps the $3$-sphere around the $S^5$. Thus the embedding is parametrized by the transverse coordinates $\psi, \th$; $\psi$ is a Killing direction and may be taken to be constant, and the appropriate solution for $\th(z)$ corresponding to massive holographic flavour is \cite{Karch:2002sh}:
\begin{equation}
	\theta(z) = \theta_c \equiv
	\begin{cases} 
		\arccos(z/z_c)	& z\leq z_c \\
		0				& z > z_c
		\end{cases}
	\end{equation}
A careful matching to the field theory shows that
\begin{equation}
	z_c = \frac{\sqrt{\lambda}}{2\pi} \frac{1}{m_F}	\label{fermion-mass}
\end{equation}
where $m_F$ is the mass of the flavour degrees of freedom \cite{Karch:2007pd, Karch:2006bv}.

Geometrically, this embedding means that the $S^3$ wrapping the $S^5$ is of maximal size ($\theta=\pi/2$) at $z=0$ on the boundary, and the $D7$-brane vanishes ($\theta =0$) at the critical value $z=z_c$. For $z>z_c$ the $D7$-brane has no effect; this is dual to the fact that at energies smaller than the mass gap the flavour degrees of freedom can no longer be excited. If we set the mass to zero $\theta$ is constant and the theory is conformal. 

If we now substitute the on-shell angle $\theta = \theta_c$ back into the DBI action and expand to quadratic order in $B_2$ and $F_2$ we obtain the following quadratic action for the fluctuations of the DBI gauge field: 		
\begin{equation}
N_f \; S_{\text{DBI}} = -\frac{N_c^2}{8 \pi^2 R^3} \int{\frac{1}{2} \kappa^2 \mu f(z)\left(B_2 + \frac{2 \pi R^2}{\sqrt{\lambda}} F_2 \right)^2}
\end{equation}
Some details about this computation are given in Appendix \ref{DBI appendix}. 
\subsection{Examples in lower dimensions}
The action \eqref{bulkac} has several interesting features, arising from the interplay of the higher form symmetry with the baryon number symmetry. To the best of our knowledge these have not yet been fully explained in the literature, and we will unpack these below. It is first helpful to orient ourselves with some more familiar examples in lower dimension. 

Let us begin with the following action for a Goldstone mode in three dimensions:
\begin{equation}
S_{1} = -\int d^3x \; \frac{v^2}{2} |d\theta|^2 \label{goldstone0form} 
\end{equation}
Clearly this has a single degree of freedom, which is gapless. The situation is however very different if we consider gauging this scalar Goldstone with a 1-form gauge field $a_{\mu}$, resulting in the following action
\begin{equation}
S_{2} = \int d^3x \le(-\frac{1}{2g^2} \left|da\right|^2 - \frac{v^2}{2} \left|a - d\theta\right|^2\ri) \label{higgsed} 
\end{equation}
This theory is now massive; the Goldstone mode is eaten by the photon, resulting in a gapped theory with mass gap $g^2 v^2$. As turning on a small gauge coupling $g$ results in a mass gap, the weak coupling limit and the infrared limit don't commute. 

Higgsing a gauge field is one way to obtain a mass gap. Another way to give a gauge field a mass is through a Chern-Simons term \cite{Deser:1981wh}. Let us thus imagine removing the Goldstone mode and adding a second gauge field $b$ to obtain the following theory:
\begin{equation}
S_{3} = \int d^3x \le(-\frac{1}{2 g^2} \left|da\right|^2 -\frac{1}{2 v^2}\left|db\right|^2 +  a \wedge db\ri)
\end{equation}
What is the spectrum of this theory? An illuminating way to understand this is to dualise the gauge field $b$ to a scalar $\psi$. Following the standard algorithm, we find
\begin{equation}
S_{3}' = \int d^3x \le(-\frac{1}{2 g^2}\left|da\right|^2 - \frac{v^2}{2} \left|d\psi - a\right|^2 \ri) 
\end{equation}
where in terms of the original degree of freedom $db = g^2 \star(d\psi - a)$. This is essentially the same as the Higgs-ed theory studied above in \eqref{higgsed}, and is also gapped. Thus we see that adding a Chern-Simons term and Higgsing a gauge field are the same mechanism, just written in different duality frames.  

Finally, let us imagine {\it both} adding a Chern-Simons term and Higgsing, i.e. we study the following action:
\begin{equation}
S_{4} = \int d^3x \le(-\frac{1}{2 g^2} |da|^2 - \frac{1}{2 v^2} |db|^2 +  a \wedge db + \frac{v^2}{2} |a-d\theta|^2 \ri) \label{newtheta}
\end{equation}
What is the spectrum now? It is again helpful to dualise the $b$ field, after which we obtain:
\begin{equation}
S_{4}' = \int d^3x \le(-\frac{1}{2 g^2}|da|^2 - \frac{v^2}{2} |d\psi -  a|^2 - \frac{v^2}{2} |d\theta - a|^2 \ri) 
\end{equation} 
i.e. after a duality this is like ``Higgsing twice''. But we only have one photon to eat a putative Goldstone; thus there should remain one Goldstone left uneaten, which can be seen by rewriting the action to be:
\begin{equation}
S_{4}'' = \int d^3x \le(-\frac{1}{2g^2}|da|^2 - \frac{v^2}{4}|d\psi + d\th - 2 a|^2 - \frac{v^2}{4}|d\psi-d\th|^2\ri)
\end{equation}
Thus the gauge-charged combination $\psi + \th$ is eaten, and forms part of a massive photon; however the gauge-invariant combination $\psi - \th$ remains uneaten, and is a gapless mode in the spectrum. The general lesson is that if we try to give a gauge field a mass {\it twice}, both by Higgsing and by adding a Chern-Simons term, then we find that a gapless mode survives. We could also have chosen to dualise the scalar $\th$ in \eqref{newtheta} into a 1-form gauge field; in this case the gapless mode would have appeared to be a gauge field and not a scalar, but this gauge field would be related to $\psi - \th$ by the usual Abelian duality. 

What does all of this have to do with AdS/CFT? In most discussions of holographic flavour in AdS/CFT, one works in the probe limit, considering the limit $\mu \sim \frac{N_f}{N_c}$ to $0$ and studying the fluctuations of the DBI worldvolume gauge field $A_1$, which then decouples from the other fields, and whose action takes the form $S \sim \int (dA_1)^2$. This is the action of a massless photon and is analogous to a higher-form version of \eqref{goldstone0form}. The field theory dual of this massless gauge field is the baryon number current. 

However, if we consider the complete action \eqref{bulkac}, we see that this is actually somewhat dangerous; in fact the gauge field does not appear by itself but rather in the gauge-invariant combination $B_2 + \frac{2 \pi R^2}{\sqrt{\lam}}F_2$. At first glance, this appears somewhat problematic, as the action contains the following terms:
\begin{equation}
	\frac{N_c^2}{8 \pi^2 R^3}\int\left[- \frac{1}{2} (dB_2)^2  - \frac{1}{2} \kappa^2 \mu \left(B_2 + \frac{2 \pi R^2}{\sqrt{\lambda}} F_2 \right)^2 + \cdots \right] 
\end{equation}
(where for simplicity we consider a case where $f(z)$ is constant). Compare this to \eqref{higgsed}; it actually now appears that the field $A_1$ has been {\it eaten} by the higher form gauge field $B_2$, in a higher-form analogue of the Higgs mechanism. This suggests that the theory should have only massive modes, with the mass scaling like $\ka^2\mu$. In a sense, the infrared limit no longer commutes with the probe limit. However, this is clearly a nonsensical result: the dual field theory should still have a baryon number current, which should be dual to a bulk gauge field that is exactly massless to all orders in $\frac{N_f}{N_c}$. 

The resolution comes in studying the full action, which contains an extra field $C_2$ which has its own kinetic term as well as a mixed Chern-Simons term coupling it to $B_2$. The action now appears more like a higher form version of \eqref{newtheta}, which indeed {\it does} support a gapless mode, though it is not apparent at first glance. In this work we will unpack the analogous mechanism in the AdS/CFT context; we will indeed find that the bulk action always supports a gapless mode. For a certain set of boundary conditions (those which are dual to the $SU(N)$ gauge theory), this mode is dual to the baryon number current. For a different set of boundary conditions (those which are dual to the $U(N)$ gauge theory), this mode is dual to the 2-form current for magnetic flux.  

The fact that the existence of this gapless mode depends crucially on the interplay between the Chern-Simons and DBI terms is dual to the fact that in the field theory the baryon number current is intertwined in some sense with the 1-form center $\mathbb{Z}_{N}$ symmetry of the pure gauge theory. 

\section{\texorpdfstring{$SU(N)$}{SU(N)} gauge theory} \label{sec:SU} 
The bulk may be understood in various duality frames. We begin by studying it in a frame which is useful when the dual field theory is the $SU(N)$ gauge theory coupled to fundamental flavour. 
\subsection{Bulk action}
To begin, it is helpful to Poincar\'e dualise the R-R form $C_2$ to a $1$-form $\tilde{C}_1$ in the usual way. The procedure is explained in for example Appendix B.4 of \cite{Polchinski:1998rr}; applying the algorithm we find:
\begin{equation}
	\tilde{G}_2 = \frac{\lambda}{4 \pi N_c} \star G_3 - \kappa B_2 \label{tildeCdef} 
\end{equation}
where $\tilde{G}_2 = d\tilde{C}_1$. Substituting into the action now gives a different presentation of the same theory. 
\begin{equation}
	S = - \frac{N_c^2}{8 \pi^2 R^3}\int{\left[\frac{1}{2}H_3^2 + \frac{1}{2}\kappa^2 \left(\left(B_2 + \frac{1}{\kappa} \tilde{G}_2\right)^2 + \mu f(z)\left(B_2 + \frac{2 \pi R^2}{\sqrt{\lambda}} F_2 \right)^2 \right)\right]}
\end{equation}
Observe now that from the duality relation (\ref{tildeCdef}), $\tilde{C}_1$ inherits the gauge-transformation under the 1-form gauge symmetry:
\begin{equation}
\delta \tilde{C}_1 = - \kappa \left(\Xi^{(B)}_1 + d \xi^{(C)}\right)
\end{equation}
which ensures that the action is indeed still gauge-invariant. Here $\Xi^{(B)}$ is the original 1-form gauge transformation of $B_2$, whereas $\xi^{(C)}$ is an emergent 0-form gauge transformation that exists only in this duality frame. In a sense it is the magnetic dual of the 1-form gauge transformation of $C_2$, which has been dualised away. 

We can diagonalise the action in this duality frame to better understand its spectrum. After diagonalising, it will also be easier to fix a gauge and solve the equations of motion. To this end, it is convenient to define the function $h(z)$ by
\begin{equation}
	h(z) = \frac{1}{1+\mu f(z)}		\label{h def}
\end{equation}
so that we can define the new $1$-form fields
\begin{subequations}
	\begin{align}
		\eta_1 &= \frac{1}{\kappa} \; \tilde{C}_1 - \frac{2 \pi R^2}{\sqrt{\lambda}} \; A_1 \label{etadef}	\\
		\tau_1 &= \frac{2 \pi R^2}{\sqrt{\lambda}} \; A_1 + h \; \eta_1 \label{taudef}
	\end{align}
\end{subequations}
and their respective field strengths
\begin{subequations}
	\begin{align}
	Y_2 &= d \eta_1		\label{Y2-def}		\\	
	T_2 &= d\tau_1		\label{T2-def}
	\end{align}
\end{subequations}
These linear combinations inherit the following gauge-transformations
\begin{subequations}
	\begin{align}
	\delta \eta_1 &= d\xi		\\
	\delta \tau_1 &= -\Xi_1^{(B)} + h \; d\xi
	\end{align}
\end{subequations}
where $\xi = \xi^{(A)} - \xi^{(C)}$. Observe that $\eta_1$ has the same gauge transformation as an ordinary free Maxwell field. When the field theory is gapless, $\eta_1$ is precisely a massless gauge field in the bulk, so is the holographic dual of a $0$-form symmetry in the field theory. This $0$-form symmetry corresponds to baryon number conservation. 

The diagonalised action can now be written cleanly as
\begin{equation}
	S = -\frac{N_c^2}{8 \pi^2 R^3}\int\left\{\frac{1}{2} H_3^2 + \frac{1}{2} \kappa^2 \left[(1-h) Y_2^2 + h^{-1} \left(B_2 + T_2 + \eta_1 \wedge dh\right)^2 \right] \right\}
\end{equation}
Gauge-invariance is easy to check: $H_3$, $Y_2$ and $B_2 + T_2 + \eta_1 \wedge dh$ are each individually gauge-invariant quantities. 

This action is somewhat complicated, as it is dual to an RG flow captured holographically by the profile of the function $f(z)$. To understand the symmetry structure, it is helpful to consider the limit of zero fermion mass $m_F \to 0$. We now have $z_c \to \infty$, and so $f(z) = 1$ for  all $z$. This gives $h(z) = (1+\mu)^{-1}$.  

We then find:
\begin{equation}
	\label{noFermionMass}
	S \to -\frac{N_c^2}{8 \pi^2 R^3}\int{\left[\frac{1}{2} H_3^2 +  \frac{1}{2}\kappa^2 (1+\mu) (B_2 + d\tau_1) ^2 + \frac{1}{2} \kappa^2 \left(\frac{\mu}{1+\mu} Y_2^2\right) \right]}
\end{equation}
We can partially gauge-fix to an analogue of unitary gauge in which we set $T_2 = 0$. This describes a 2-form gauge field $B_2$ which has been Higgs-ed by the 1-form $\tau_1$; the resulting dynamical bulk field is massive. It also has a precisely massless 1-form gauge field $\eta_1$, as anticipated above. This is dual to the baryon number current. Note that this structure arose out of the interplay between the Chern-Simons term and the flavour terms; this is dual to the interplay between the $U(1)$ baryon number current and the $\mathbb{Z}_N$ center symmetry of the field theory. In the remainder of this section we further describe some universal aspects of this interplay. 

Up to boundary terms, the variation of the action is 
\begin{equation}
	\begin{split}
	\delta S
	= - \frac{N_c^2}{8 \pi^2 R^3} \int \bigg \{
	& \delta B_2 \wedge \star \left[\star d \star dB_2 + \kappa^2 h^{-1} (B_2 + T_2 + \eta_1 \wedge dh) \right]	\\
	& - \kappa^2 \; \delta \eta_1 \wedge \star \left[\star d((1-h) \star d \eta_1) -h^{-1} \star (dh \wedge \star (B_2 + T_2 + \eta_1 \wedge dh)) \right]	\\
	& + \kappa^2 \; \delta \tau_1 \wedge d \left[h^{-1} \star (B_2 + T_2 + \eta_1 \wedge dh) \right] \bigg \}
	\end{split}
\end{equation}
Hence the equations of motion are\footnote{The $\tau_1$ equation is redundant since it follows by taking the exterior derivative of the $B_2$ equation.}
\begin{subequations}
	\label{1-form-eom}
	\begin{align}
		\star \; d \star d B_2 +  \kappa^2 h^{-1} (B_2 + T_2 + h' \eta_1 \wedge dz) &= 0		\\
		(1-h) \star d \star d \eta_1 + h^{-1} h' \star [dz \wedge \star (-h \; d\eta_1 + B_2 + T_2 + h' \eta_1 \wedge dz)] &= 0
	\end{align}
\end{subequations}
where $h' \equiv \frac{dh}{dz}$.

The spectrum of fields thus consists of a massive $2$-form gauge field $B_2$, a massless $1$-form gauge field $\eta_1$ which is dual to the baryon number current, and a $1$-form gauge field $\tau_1$ which appears only through its field strength $T_2$. $T_2$ is of less physical importance since it can be easily gauged away; in a sense it simply provides the longitudinal degrees of freedom of the massive tensor $B_2$. 

If we are studying the $SU(N)$ gauge theory coupled to fundamental flavour, it is important to note that the appropriate boundary conditions at the AdS boundary are those where we hold fixed the boundary value of $\eta_1$; this guarantees that we obtain a conserved $0$-form baryon number current $j^{\mu}_b$ in the boundary theory. As we will see, this will be different when we study the $U(N)$ gauge theory. 

\subsection{Charged operators}
We now describe the bulk object that is charged under the baryon number symmetry. 

\subsubsection{Baryon vertices in pure \texorpdfstring{$SU(N)$}{SU(N)} gauge theory}
Let us first review the conventional case with no flavour branes \cite{Witten:1998xy}. There we set $N_f \to 0$, and we find simply:
\begin{equation}
	S = -\frac{N_c^2}{8 \pi^2 R^3}\int{\left[\frac{1}{2} H_3^2 + \frac{1}{2} \kappa^2 (B_2 + T_2)^2 \right]}
\end{equation}
With no flavour branes the DBI gauge field $A_1$ does not exist, and from \eqref{h def} and \eqref{taudef} we see that when $h = 1$ we have simply $\tau_1 = \kappa^{-1} \tilde{C}_1$, i.e. $\tau_1$ is directly the magnetic dual of the RR 2-form. 

We will now revisit this action from the point of view of symmetry. Note that under the 1-form gauge transformation of $B_2$, $\tau_1$ must also transform:
\begin{equation}
B_2 \to B_2 + d\Xi_{1} \qquad \tau_1 \to \tau_1 - \Xi_{1} \label{xitrans} 
\end{equation}
We now study the bulk objects that are charged under these gauge symmetries. We have fundamental string worldsheets, which couple minimally to $B_2$ as
\begin{equation}
\frac{1}{2\pi l_{s}^2} \int_{\sM} B_2 
\end{equation}
We turn now to $\tau_1$; as $\tau_1$ is a usual 1-form gauge field in the bulk, it couples naturally to one-dimensional particle worldlines in AdS$_5$. What are these objects? 

From the definition of the duality relationship \eqref{tildeCdef}, we can see that these objects couple magnetically to the RR 2-form $C_2$. In the ten-dimensional picture, these are thus $D5$-branes. Of their six dimensional worldvolume, five of them are wrapped on the $S^5$, and the remaining one dimension traces a worldline on AdS$_5$. By using the normalisations in Appendix \ref{app:CS} one can verify that a single such $D5$-brane couples to $\tau_1$ as
\begin{equation}
\frac{N_c}{2\pi l_s^2} \int_{L} \tau_1
\end{equation}
Note however that this coupling alone is not invariant under the 1-form gauge transformation \eqref{xitrans}: indeed we see that the the one-dimensional worldline $L$ can exist only as the {\it boundary} of $N_c$ string worldsheets, i.e. the combined coupling
\begin{equation}
N_c \le(\frac{1}{2\pi l_{s}^2} \int_{\sM} B_2 \ri) + \frac{N_c}{2\pi l_s^2} \int_{\p\sM} \tau_1
\end{equation} 
is invariant. This fact -- that the wrapped $D5$-brane is the endpoint of $N_c$ fundamental strings and thus acts as a baryon vertex in the dual field theory -- can also be understood directly from the original Chern-Simons coupling \cite{Witten:1998xy,Aharony:1998qu,Maldacena:2001ss,Witten:1998wy}. Here we simply restate it in an alternative (bulk) duality frame. 

It is now instructive to imagine the bulk worldline intersecting the AdS boundary at a point.  Each of the $N_c$ string worldsheets will also intersect the boundary as a series of curves ending at the same point. Holographically, the combined object is a non-dynamical baryon vertex serving as the endpoint of $N_c$ Wilson lines in the fundamental representation. It is clear that the baryon vertex, being tied to $N_c$ Wilson lines, is not a local operator in the field theory; the bulk dual of this statement is that it does not correspond to a free particle worldline but rather only as the boundary of $N_c$ F-strings. 
\begin{figure}[h!]
\begin{center}
\includegraphics[scale=0.4]{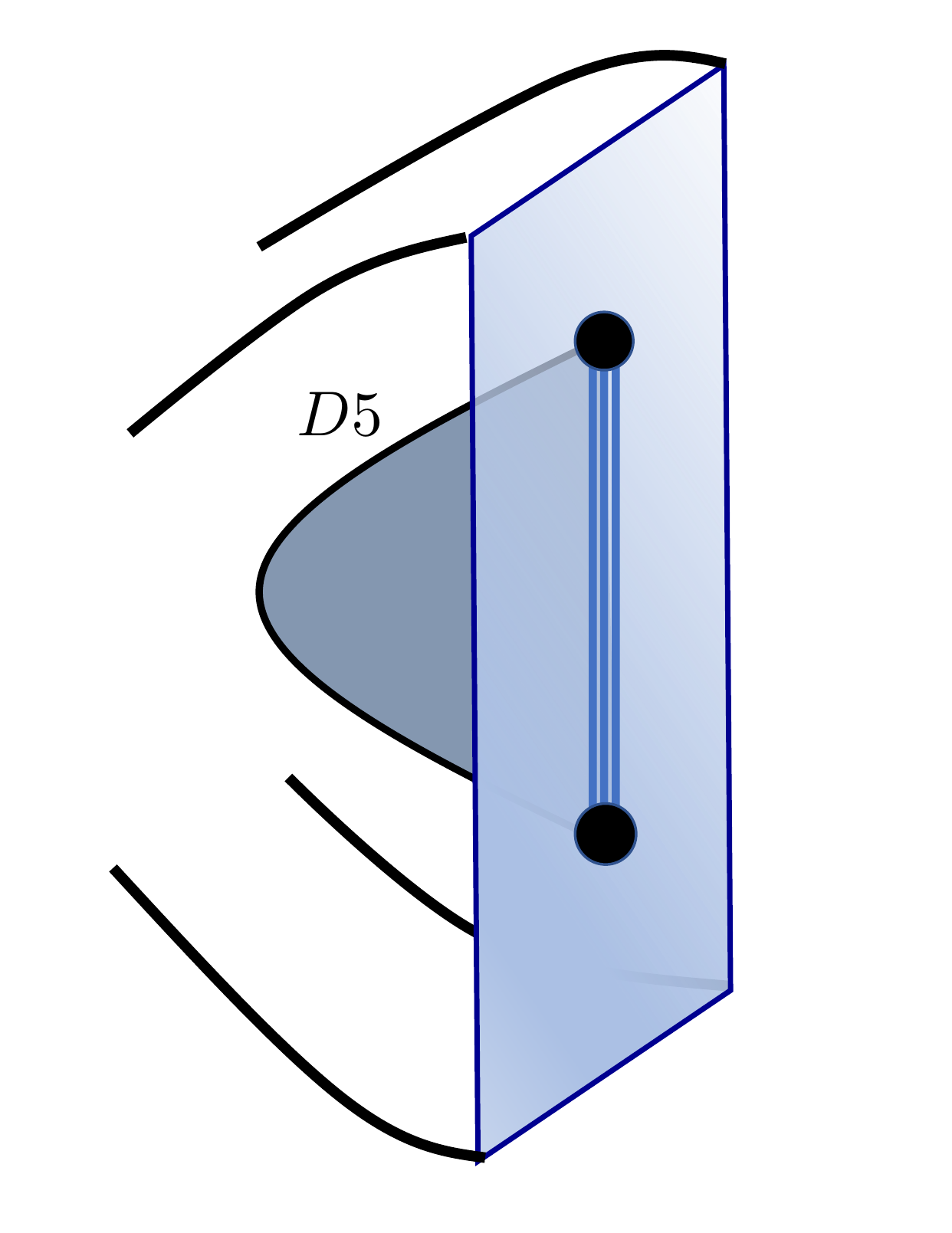}
\caption{\label{fig:hangingD5}
The hanging $D5$-brane forms 1-dimensional worldline in the bulk; it is the boundary of $N_c$ F-string worldsheets which also intersect the AdS boundary. At their intersection with the boundary they define the insertion of fundamental Wilson lines in the dual CFT.}
\end{center}
\end{figure}

\subsubsection{Baryon operator in theory with dynamical flavour}
We now restore the flavour branes, i.e. we return to \eqref{noFermionMass}. In the bulk, we now have a new massless field $\eta_1$, which we understand is dual to the baryon number current in the field theory $U(1)_B$. In the dual field theory, we now expect the existence of local baryon operators that carry charge $N_c$ (in units of the baryon charge of the fundamental gauge-charged fields). 

What is the bulk dual of this operator? Consider a general particle-like object in the bulk that couples to both $\eta_1$ and $\tau_1$, i.e.
\begin{equation}
\int_{L} \le(q_{\eta} \eta_1 + q_{\tau} \tau_1\ri)
\end{equation}
As argued above, any coupling to $\tau_1$ will necessarily mean that the particle has strings attached, in order to ensure gauge-invariance. Let us consider an object which has $q_{\tau} = 0$. As $\eta_1$ does not transform under the 1-form gauge transformation \eqref{xitrans}, this coupling is entirely gauge-invariant on its own. Thus a particle in the bulk that couples in this way is dual to a local boundary operator that carries baryon charge. From field theory considerations, we understand that this object should be related to a bound state of a $D5$-brane and F-strings in some manner. 

The presence of the new field $\eta_1$ lets the $D5$-brane exist as an independent object that is untethered to any strings. To see this more explicitly, we can express this coupling in terms of the original fields $\tilde{C}_1$ and $A_1$; we find that the unit quantized $D5$-brane couples as
\begin{equation}
q_{\eta} \int_{L} \eta_1 = \frac{N_c}{2\pi l_{s}^2} \int_{L} \le(\frac{2 \pi R^2}{\sqrt{\lam}} A_1 - \kappa \; \tilde{C}_1 \ri)
\end{equation}
where we have used the quantized coupling to $\tilde{C}_1$ worked out above, and where the coupling to $A_1$ is correlated with that of $\tilde{C}_1$ by the condition that $q_{\tau} = 0$. This can be compared to the coupling of a single F-string ending on the $D7$-brane:
\begin{equation}
\frac{1}{2\pi l_s^2}\le(\int_{ws} B_2 + \int_{\p ws} \frac{2 \pi R_2}{\sqrt{\lam}} A_1\ri)
\end{equation}
In other words, the coupling to $A_1$ is as $N_c$ F-strings. Microscopically one can actually imagine that the $D5$-brane is connected by very small strings to the flavour $D7$-brane, where the string charge is now carried by the $A_1$ field living on the brane. The resulting composite object is the particle-like excitation that we describe above. Related work in different holographic models to directly construct bulk objects carrying baryon number can be found in \cite{Hata:2007mb,Seki:2008mu,Sfetsos:2008yr}; see in particular \cite{Sfetsos:2008yr}. We stress that our construction makes no real statements about the dynamics of the internal structure, and simply shows how their charges are captured in the low-energy description.

\section{\texorpdfstring{$U(N)$}{U(N)} gauge theory} \label{sec:U} 
We would now like to understand the theory with the boundary conditions that are appropriate to having a $U(N)$ gauge theory dual. We now expect to obtain a 2-form conserved current on the boundary; it is thus appropriate to study the bulk in a different duality frame. 

\subsection{Bulk action}
We can Poincar\'e dualise the field $\tau_1$ by integrating out its field strength $T_2$ in the usual way. This yields a $2$-form $\mathcal{A}_2$ with field strength $\mathcal{F}_{3} = d \mathcal{A}_2$ given by: 
\begin{equation}
	\mathcal{F}_3 = \kappa^2 h^{-1} \star \left(B_2 + T_2 + \eta_1 \wedge dh \right) \label{f3def} 
\end{equation}
Substituting into the action and integrating by parts gives
\begin{equation}
S = -\frac{N_c^2}{8 \pi^2 R^3}\int{\left[\frac{1}{2} H_3^2 + \frac{1}{2}\kappa^2 (1-h)Y_2^2 + \frac{1}{2} \kappa^{-2} h \; \mathcal{F}_3^2 + B_2 \wedge \mathcal{F}_3 - Y_2 \wedge \mathcal{A}_2 \wedge dh\right]}
\end{equation}
Note the last term in the action where $\sA_2$ appears explicitly; this arises from an integration by parts so that the action depends on $Y_2 = d\eta_1$ and not $\eta_1$ explicitly. As a result we can now dualise $\eta_1$ using exactly the same procedure to give a $2$-form $\mathcal{P}_2$ whose field strength $\mathcal{Q}_3 = d\mathcal{P}_2$ is given by 
\begin{equation}
	\mathcal{Q}_3 + \mathcal{A}_2 \wedge dh  = \kappa^2 (1-h) \star Y_2 \label{Q3def} 
\end{equation}
(Note that $\sP_2$ can be thought of as -- modulo mixing with other fields -- the electric-magnetic dual of $\eta_1$, i.e. the bulk field dual to the baryon number current). Substituting this back into the action then gives:	 
\begin{equation}
    S = -\frac{N_c^2}{8 \pi^2 R^3}\int{\left[\frac{1}{2} H_3^2 + B_2 \wedge \mathcal{F}_3 + \frac{1}{2} \kappa^{-2} \left( (1-h)^{-1} (\mathcal{Q}_3 + \mathcal{A}_2 \wedge dh)^2 + h \mathcal{F}_3^2\right) \right]}
\end{equation}
Note that we have a gauge freedom given by
\begin{subequations}
	\begin{align}
		\delta \mathcal{A}_2 & = d\Xi_1	\\
		\delta \mathcal{P}_2 & = - \Xi_1 \wedge dh + d\Lambda_1
	\end{align}
\end{subequations}
under which the action is invariant, where $\Lambda_1$ is a new free $1$-form gauge parameter. From the perspective of the $2$-form picture, two of the equations of motion in the $1$-form picture are simply the Bianchi identities $d \mathcal{F}_3 = 0$ and $d \mathcal{Q}_3 = 0$. 

The spectrum is easiest to understand in the case where the flavour mass is zero so that $dh = 0$. We then have two coupled 2-forms $B_2$ and $\mathcal{A}_2$ which constitute a massive degree of freedom. We also have a single massless 2-form $\mathcal{P}_2$ whose dependence is only through its field strength $\mathcal{Q}_3$; this massless bulk is dual to the only conserved $2$-form current $J^{\mu\nu} = J_b^{\mu\nu}$, identified in \eqref{1formcurrs}. 

The equations of motion in the $2$-form picture are
\begin{subequations}
	\begin{align}
	d \star [(1-h)^{-1}(\mathcal{Q}_3 + \mathcal{A}_2 \wedge dh)] & = 0	\\
	d \star H_3 - \mathcal{F}_3 & = 0		\\
	d \star (h \mathcal{F}_3) + \kappa^2 H_3 - (1-h)^{-1} dh \wedge \star (\mathcal{Q}_3 + \mathcal{A}_2 \wedge dh) & = 0
	\end{align}
\end{subequations}
From the perspective of the $1$-form picture, two of these equations give the Bianchi identities $d Y_2 = 0$ and $d T_2 = 0$. The third equation is the same in both pictures.

Thus, to obtain the bulk dual to the $U(N)$ gauge theory coupled to flavour, we should use AdS/CFT boundary conditions where we hold fixed the boundary value of the 2-form field $\sP$. The usual rules of AdS/CFT will then guarantee that in the dual field theory, we will obtain a 2-form conserved current $J$, as expected. 

We note that the $U(N)$ theory seems to contains one extra parameter as compared to the $SU(N)$ theory; as explained around \eqref{deformLag2}, the coupling constant $g_1$ associated to the ``$U(1)$ factor'' seems to be an extra parameter that can be tuned. In a universal sense this can be understood as a double-trace coupling associated to the 2-form current $J$. When there are flavor degrees of freedom present this coupling is expected to run logarithmically, becoming strong in the UV. Thus, due to dimensional transmutation the extra data that needs to be provided is not a dimensionless coupling but rather the energy scale for the Landau pole at which this coupling becomes strong. As explained in \cite{Hofman:2017vwr,Faulkner:2012gt}, the boundary conditions for a massless 2-form field such as $\sP$ in AdS$_5$ indeed require one to specify such a scale. We will see this explicitly when solving the bulk equations of motion in later sections. 

We provide a few more details; as usual, $J$ is obtained by taking a functional derivative of the bulk on-shell action with respect to the boundary value of $\mathcal{P}_2$. If we set 
\begin{equation}
	\lim_{z\to 0}{\mathcal{P}_2} = p_2
\end{equation}
and use the equation of motion then we obtain
\begin{align*}
	J^{\mu\nu} = \frac{\delta S_{\text{on-shell}}}{\delta p_{\mu\nu}}
	& = \frac{\mathcal{N}}{2 \kappa^2} \frac{\delta}{\delta p_{\mu\nu}} \int{d\left[(1-h)^{-1}\mathcal{P}_2 \wedge \star (\mathcal{Q}_3 + \mathcal{A}_2 \wedge dh)\right]}		\\
	& = \frac{\mathcal{N}}{2} \frac{\delta}{\delta p_{\mu\nu}} \int{d(\mathcal{P}_2 \wedge Y_2)}
\end{align*}
From here we can conclude that the 2-form symmetry current is: 
\begin{equation}
J^{\mu\nu} = \lim_{z \to 0} \frac{\sN}{2} \left(\star_{4} Y_2\right)^{z \mu \nu}
\end{equation}
where the normalisation is given by $\mathcal{N} = \frac{N_c^2}{8 \pi^2 R^3}$. 

\subsection{Charged line operator}
We would now like to understand the bulk operators that are charged under the $2$-form gauge field $\sP_{2}$. In the field theory, these are dual to line operators that are charged under the corresponding $1$-form symmetry. In this subsection only we will work only to first order in $\mu$ to simplify the formulas. We begin by tracing back through the chain of dualities; from \eqref{etadef} to \eqref{taudef}, in the small $\mu$ limit we find:
\begin{equation}
\tau_1 = \frac{1}{\ka} \tilde{C}_1 + 2\pi \ell_s^2 \mu A_1 \qquad
\eta_1 = \frac{1}{\ka} \tilde{C}_1 - 2\pi \ell_s^2 A_1 \label{allcoups} 
\end{equation}
Furthermore, in the same limit we find
\begin{equation}
d \sP_{2} = \ka^2 \mu \star d\eta_1 \qquad d\sA_{2} = \ka^2 \le(B_2 + d\tau_1\ri)	
\end{equation}
An object which couples minimally to $\sP_2$ is one that appears on the right hand side of the equation of motion $d \star d\sP_2 = 0$; we thus need to find bulk objects that couple {\it magnetically} to the fields $\tilde{C}_1$ and $A_1$. As $\tilde{C}_1$ is the magnetic dual of the RR 2-form $C_2$, the object coupling magnetically to it is simply a D1-string. In the Appendix we work out the normalisation of this coupling in our conventions to show that 
\begin{equation}
\frac{1}{\ka} \int_{S^2} d\tilde{C}_1 = \frac{(2\pi \ell_s)^2}{N_c} \label{D1coupling} 
\end{equation}
where here the $S_2$ wraps a $D1$-string that is hanging down into AdS$_5$. 

The object which couples magnetically to the DBI worldvolume gauge field $A_1$ is somewhat more interesting. We will call this object the DBI monopole. In this section we will work in the case where $N_f = 1$; the situation for generic $N_f$ is more interesting still and we will touch on it briefly later. A similar problem was discussed in \cite{Iqbal:2014cga} in a lower dimensional construction, and we may take over the same ideas. The desired magnetically charged object turns out to be a wrapped $D5$-brane that {\it ends} on the $D7$ flavour brane. To be more precise, recall from the earlier sections that the $D7$ flavour brane wraps an $S^3 \subset S^5$:
\begin{equation}
d\Omega_5^2 = d\theta^2 + \cos^2{\theta} \; d\psi^2 + \sin^2{\theta} \; d\Omega_3^2 \label{S5repeat} 
\end{equation}
where the $S^3$ is spanned by the coordinates $\Omega_3$. The $D7$-brane does not extend in $\theta$: more precisely, for each value of the radial coordinate $z$, the $D7$-brane sits at a single $\theta(z)$. In the conformal case, it lives at $\theta_{D7} = \frac{\pi}{2}$ for all $z$, whereas in the non-conformal case $\theta_{D7}$ interpolates from $\frac{\pi}{2}$ at the UV boundary to $0$ in the interior. 

\begin{figure}[h!]
\begin{center}
\includegraphics[scale=0.6]{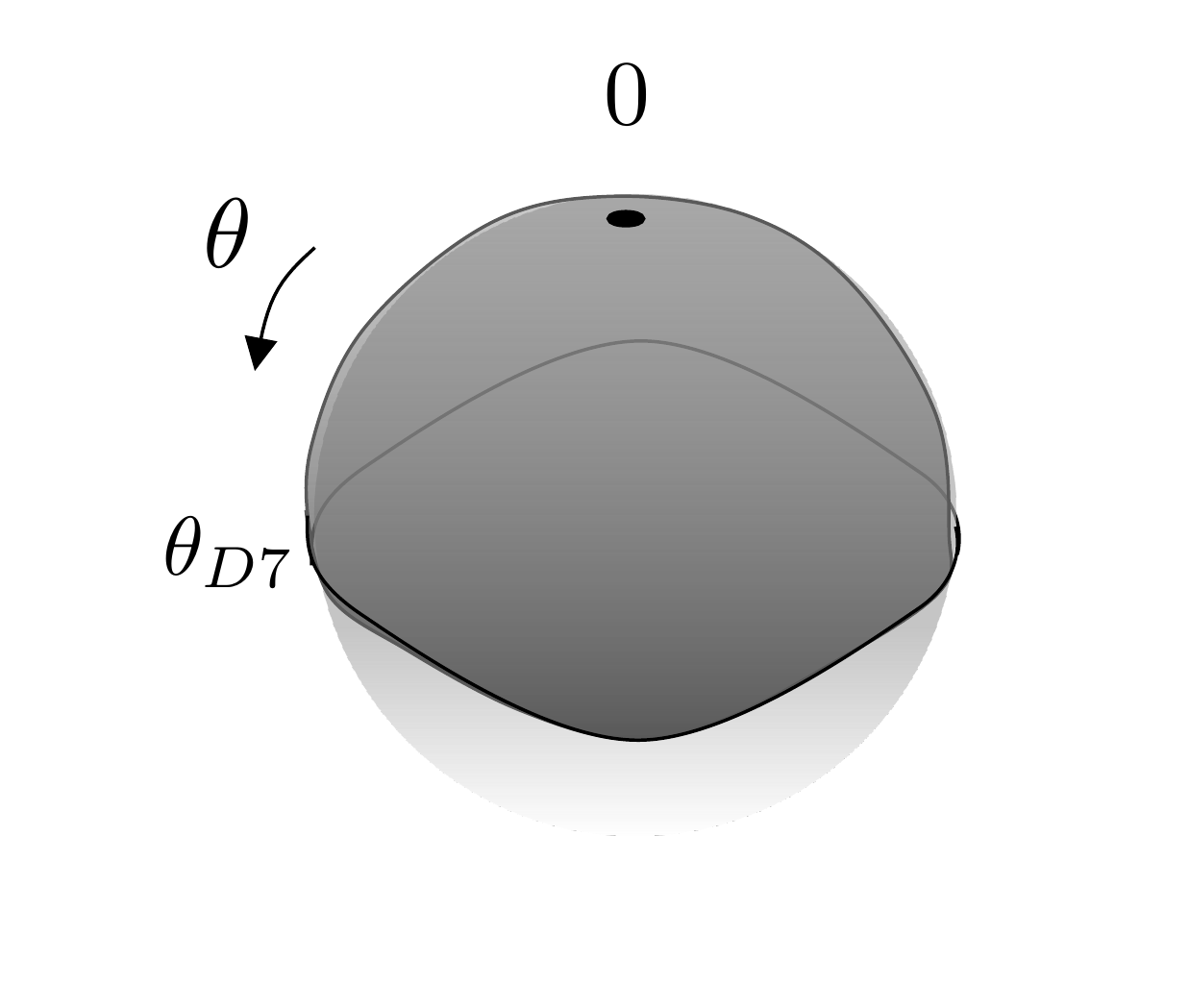}
\caption{\label{fig:halfcap}
The $D5$-brane wrapping half of the $S^{4} \subset S^{5}$ formed by $(\th, \Om_{3})$, ending on the $D7$-brane which lives at $\theta = \theta_{D7}$. The remaining 2 coordinates of the $D5$-brane worldvolume form a two-dimensional string worldsheet in the bulk. }
\end{center}
\end{figure}

In contrast, consider a $D5$-brane that wraps this $S^3$ and {\it ends} on the $D7$-brane. The $D5$-brane extends in $\theta$ from $\theta = 0$ to the $\theta_{D7}$ coordinate of the $D7$-brane, as shown in Figure \ref{fig:halfcap}. It sits at a particular value of $\psi$; as $\psi$ is a Killing direction this choice is arbitrary. 

The boundary of the $D5$-brane is a five-dimensional manifold; three of these dimensions are compact and form the $S^3$, and the remaining two dimensions define a two dimensional manifold $\sM_2$ which extends into the bulk of AdS$_5$. As is well known \cite{Strominger:1995ac}, the boundary of this $D5$-brane appears magnetically charged to the DBI gauge field $A_1$ living on the $D7$-brane worldvolume. Hence the wrapped $D5$-brane is the DBI monopole that we seek. 

In the Appendix, we work out the coupling of this brane and show that the coupling to one such wrapped brane is
\begin{equation}
2\pi \ell_s^2 \int_{S^2} F_2 = (2\pi \ell_s)^2 \label{disorderF} 
\end{equation}
where $F_2 = dA_1$ and the $S^2$ surrounds $\sM_2$ in AdS$_5$. By comparing this to \eqref{D1coupling} and \eqref{allcoups}, we see that the $D1$-brane couples to $\sP$ with $1/N_{c}$ the charge of the DBI monopole. We may write an effective coupling to the $\sP$ field for both of these objects:
\begin{equation}
S = q_{D1} \int_{D1} \sP + q_{D5} \int_{D5} \sP
\end{equation}
The overall normalisation of $q_{D1}$ and $q_{D5}$ depends on the (arbitrary) convention chosen to normalize $\sP$ in our action, but their ratio is fixed on topological grounds to be $N_{c}^{-1}$.
\begin{equation}
\frac{q_{D1}}{q_{D5}} = \frac{1}{N_c} \label{ratio}
\end{equation}

Let us now turn to an understanding of this from the dual field theory. The intersection of the $D1$ string with the AdS boundary defines a t'Hooft line in the $SU(N)$ gauge theory sector. Similarly the wrapped $D5$-brane defines a t'Hooft line for the $U(1)$ gauge theory sector; the simplest way to see this is to note that when evaluated at the boundary, \eqref{disorderF} is precisely the definition of a t'Hooft line. It has been previously noted (see e.g. Appendix C of \cite{Hofman:2017vwr}) that from the point of view of the $U(1)$ magnetic 1-form current, the charge of a non-Abelian t'Hooft line has $U(1)$ charge of 1/$N_{c}$-th the Dirac monopole, consistent with \eqref{ratio}. 

Let us now understand the dynamics of symmetry breaking. Consider the wrapped DBI monopole such that it intersects the AdS boundary on a 1d curve $C$. This defines the insertion of a line operator into the field theory $\langle W(C) \rangle$, and as usual from the rules of AdS/CFT we should compute:
\begin{equation}
\langle W(C) \rangle \sim \exp\le(-S_{D5}[C]\ri)
\end{equation} 
with $S_{D5}[C]$ the on-shell action of the wrapped $D5$-brane. We now seek to understand the dependence of this answer on the curve $C$; if it depends only locally on the data of the curve $C$ (e.g. as a perimeter law) then the symmetry is spontaneously broken. If it depends non-locally -- e.g. as an area law, or more generally in any way that cannot be locally determined by the curve, then the symmetry is unbroken.

\begin{figure}[h!]
\begin{center}
\includegraphics[scale=0.4]{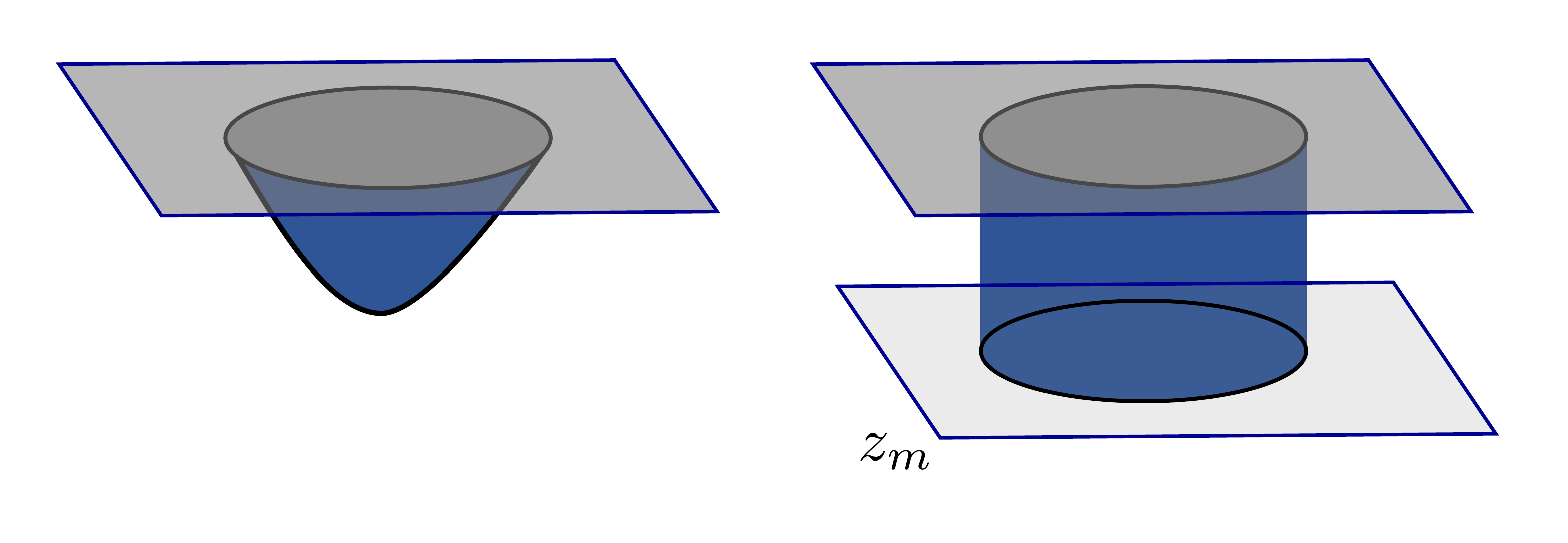}
\caption{\label{fig:areaperimeter}
Two distinct topologies that are possible for the DBI monopole. On the left is the situation when the flavour sector is gapless; the $D7$-brane then has no boundary, and the DBI monopoles hangs down into the bulk with a disc topology, whose action depends non-locally on the data describing the boundary curve. On the right, when the flavour brane ends, the $D5$-brane is also allowed to end, permitting a cylinder topology. The action of this configuration depends only on the perimeter of the boundary curve.}
\end{center}
\end{figure}

The precise holographic arguments are a higher-form analogue of the arguments presented in \cite{Iqbal:2014cga}. Consider first the case where the mass of the flavour degrees of freedom is zero, i.e.  $z_c \to \infty$. In this case the $S^3$ factor of the $D7$-brane remains the same size everywhere in the bulk, i.e. it is independent of $z$. As the brane always hangs down into the bulk, this defines a minimal area problem, essentially the same as in the usual studies of Wilson lines from holography \cite{Maldacena:1998im}. It is clear from the geometry that the on-shell action will always depend more strongly on the curve itself than its perimeter. Thus by the previous paragraph, the symmetry is {\it unbroken}. See Figure \ref{fig:areaperimeter} for a visualisation of this geometry.

Let us now consider the case where the mass of the flavour degrees of freedom is nonzero. Then there is a value of $z_c$ at which the $D7$-brane caps off.  At this value of $z_c$ the wrapped $D5$-brane also pinches off and is allowed to smoothly end. There are now two disconnected possibilities for the topology of the hanging DBI monopole; it can form topologically a disc, or it can be topologically a cylinder which hangs straight down and ends where the brane caps off. For sufficiently large sizes of the curve, the cylinder solution will dominate. Such topologically non-trivial phase transitions are common in holography \cite{Brandhuber:1998bs,Rey:1998bq,Klebanov:2007ws}. As the surface now hangs straight down, the action will depend only on the perimeter of the curve (multiplied by a constant distance in the holographic direction), and in this phase the $U(1)$ symmetry is spontaneously broken, as expected. 

Finally, one could attempt to generalize the construction of defect operators to the case $N_F > 1$; in this case there is presumably an extra quantum number associated with which of the $N_F$ D7 branes the D5 brane ends on. It seems that a careful study of the braiding algebra of bulk operators should allow the holographic identification of the mixed symmetry of rank $\mbox{gcd}(N, N_F)$ identified by \cite{Cherman:2017tey}. We leave this for later study. 

\section{Fluctuation spectrum} \label{sec:flucspec} 

In the remainder of this paper we study only the $U(N)$ theory, with its associated 1-form symmetry associated with magnetic flux. We have argued above that if the flavour degrees of freedom are gapped, then the 1-form symmetry is spontaneously broken, as can be seen from the fact that the charged line operator exhibits a perimeter law. On general grounds, we thus expect that there exists a gapless Goldstone mode in the spectrum \cite{Hofman:2018lfz,Lake:2018dqm}. In this section we will explicitly solve the equations of motion to show the existence of this Goldstone mode. We first digress slightly to explain precisely what a Goldstone mode means in this context. 

Consider a completely general Lorentz-invariant four-dimensional quantum field theory with a conserved 2-form current $J^{\mu\nu}$. For simplicity, let us study the theory in Euclidean signature; as explained in (e.g.) \cite{Hofman:2017vwr}, the two-point function of the current in momentum space then takes the general form
\begin{equation}
	\langle J^{\mu\nu}(k) J^{\rho\sig}(-k)\rangle =  \le(-\frac{1}{k^2} \le(k^{\mu}k^{\rho} g^{\nu\sig} - k^{\nu}k^{\rho} g^{\mu\sig} - k^{\mu} k^{\sig} g^{\nu\rho} + k^{\nu} k^{\sig} g^{\mu\rho} \ri) + \le(g^{\mu\rho} g^{\nu\sig} - g^{\mu\sig} g^{\nu\rho}\ri)\ri) f_{JJ}\le(\frac{|k|}{m}\ri) \label{Jcorrdef}
\end{equation}
where here $f_{JJ}$ is a dimensionless function and $m$ is a scale. The correlation is completely determined by the function $f$. In this context, spontaneous breaking of the symmetry means that $f_{JJ}\left(\frac{|k|}{\Lam}\right)$ approaches a constant as $k \to 0$; the $k \to 0$ limit then results in a gapless mode from the inverse factors of $k^{-2}$ arising from the tensor structure. 

A simple example is given by pure 4d electrodynamics; here the 1-form symmetry is broken, and the correlator of the magnetic flux $J = \star F$ takes precisely this form with 
\begin{equation}
	f_{JJ}(k) = \frac{1}{g^2}
\end{equation}
where $g^2$ is the electromagnetic coupling. 

An example where the symmetry is {\it not} spontaneously broken is given by the holographic example studied in \cite{Hofman:2017vwr}. Here the theory in question was a simple bottom-up holographic realisation of a 1-form symmetry, and the function $f_{JJ}$ was given by
\begin{equation}
	f_{JJ}(k) = \frac{1}{g^2 \log\le(\frac{|k|}{\Lam}\ri)}
\end{equation}
where $\Lam$ is a Landau pole, i.e. a UV scale where the theory breaks down, as described in \cite{Hofman:2017vwr}. We note here that $f_{JJ}$ vanishes at $k \to 0$, and the symmetry is not spontaneously broken. A similar result is found whenever there are electrically charged degrees of freedom present that are massless. 

In this section we will explicitly solve the bulk equations of motion and compute the function $f_{JJ}$ in our theory, showing that the low-frequency limit does not vanish. We will then compare it to expectations at weak coupling.  
\subsection{Goldstone modes and numerics}

We will proceed by computing the correlation function of spatial components of $J^{ij}$ with the (Euclidean) momentum purely in the time direction. Although the Green's function of interest is better extracted in the ``$2$-form'' duality frame with the fields $\mathcal{P}_2$ and $\mathcal{A}_2$, it is easier to solve the equations of motion in the ``$1$-form'' duality frame consisting of the fields $\eta_1$ and $\tau_1$. Our strategy will be to solve the bulk equations of motion in the $1$-form frame and then exploit a simple correspondence between the frames at the UV boundary to extract the Green's function. 

For numerical convenience, we will set $\mu = 1$. As explained below \eqref{Poincare-coords}, we are working in an illustrative approximation where we capture some aspect of the backreaction of the flavor degrees of freedom on the color dynamics, while neglecting gravitational backreaction. The results below do not depend qualitatively on this choice of $\mu$, but this $\mathcal{O}(1)$ choice allows us to conveniently find  numerical solutions to the equations of motion.

\subsubsection{1-form}
We solve the equations of motion given in \eqref{1-form-eom} by partially fixing the gauge so that $T_2 = 0$. Next we Fourier transform the fields in the field theory directions and exploit Lorentz invariance to choose the momentum $k^\mu = (\omega, 0)$. 
This allows us to expand some expressions involving differential forms in terms of their components as
\begin{equation}
	\star d \star d \eta_1
	= \frac{z}{R^2} \left\{
	\left(z\eta_i'' - \eta_i' + z \omega^2 \eta_i\right) dx^i + i \omega z (\eta_t' - i \omega \eta_z) \; dz + \left[z (\eta_t'' - i \omega \eta_z') - \eta_t' + i \omega \eta_z \right] \; dt
	\right\}
\end{equation}
and
\begin{equation}
	H_3
	= \frac{1}{2} B_{ij}' \; dx^i \wedge dx^j \wedge dz + \left(B_{it}' + \frac{1}{2} i \omega B_{ij} dx^i \wedge dx^j \wedge dt - i \omega B_{iz}\right) dx^i \wedge dt \wedge dz
\end{equation}
We next note that for a general $2$-form $\Omega_2$, we have
\begin{equation}
	\star(dz \wedge \star \Omega_2)
	= - \frac{z^2}{R^2}(\Omega_{iz} \; dx^i + \Omega_{tz} \; dt)
\end{equation}
Similarly, for a general $3$-form $\Omega_3$ we have
\begin{equation}
	\star d \star \Omega_3
	= - \frac{z}{R^2}\left\{\frac{1}{2}\left((z\Omega_{ijz})'+i \omega z \Omega_{ijt}\right)dx^i \wedge dx^j + (z\Omega_{itz})'dx^i \wedge dt + i \omega z \Omega_{itz} dx^i \wedge dz\right\}
\end{equation}
In pure AdS we can also show that
\begin{equation}
	\partial_S \; \epsilon^{MNP}{}_{QR} = \frac{1}{z} \; \delta_S^z \; \epsilon^{MNP}{}_{QR}
\end{equation}
which allows us to write
\begin{equation}
	\star d \star H_3
	= -\frac{z}{R^2} \left\{\frac{1}{2}((zB_{ij}')'-\omega^2 z B_{ij})dx^i \wedge dx^j + (zH_{zit})' dx^i \wedge dt + i \omega z H_{zit} dx^i \wedge dz \right\}
\end{equation}
where 
\begin{equation}
	H_{zit} = B_{it}' - i \omega B_{iz}
\end{equation}
Now we can write the equations of motion more explicitly in components. The $B_2$ equation of motion is 
\begin{equation}
	\begin{split}
		& zh\left[\frac{1}{2}\left(zB_{ij}'\right)'dx^i \wedge dx^j + (zH_{zit})' dx^i \wedge dt + i \omega z H_{zit} dx^i \wedge dz \right]\\
		& = 16\left[(B_{iz}+h' \eta_i) \; dx^i \wedge dz + (B_{tz} + h' \eta_t) dt \wedge dz + \frac{1}{2} B_{ij} \; dx^i \wedge dx^j + B_{it} \; dx^i \wedge dt\right]
	\end{split}
\end{equation}
We are interested in the vector channel, namely the components with a single spatial index; after imposing the duality relation at the boundary this contains the information of the transverse channel of the $J^{ij}$ correlation function. 
\begin{subequations}
	\begin{align}
		zh (zH_{zit})'			& = 16 B_{it}	\\
		zh (i \omega z H_{zit}) & = 16(B_{iz}+h' \eta_i)
	\end{align}
\end{subequations}
These can be combined to give
\begin{equation}
	zh \left[\frac{z(B_{it}' + i \omega h' \eta_i)}{16 - \omega^2 z^2 h} \right]' - B_{it} = 0
\end{equation}
where we eliminated $B_{iz}$ using
\begin{equation}
	B_{iz} = \frac{i \omega z^2 h B_{it}' - 16h' \eta_i}{16 - \omega^2 z^2 h}
\end{equation}
We also have the $\eta_1$ equation 
\begin{align}
	&(1-h)\left[\left(z\eta_i'' - \eta_i' + z \omega^2 \eta_i\right) dx^i + i \omega z (\eta_t' - i \omega \eta_z) dz + (z (\eta_t'' - i \omega \eta_z') - \eta_t' + i \omega \eta_z)dt\right]		\\
	&- zh^{-1}h'[(h \eta_i' + h' \eta_i + B_{iz})dx^i + (h \eta_t' - i \omega h \eta_z + B_{tz} + h' \eta_t)dt] = 0
\end{align}
We are most interested in the vector channel equation
\begin{equation}
	\label{eta-vector}
	(1-h) \left(\eta_i'' - \frac{1}{z} \eta_i' + \omega^2 \eta_i \right) - h^{-1}h' \left(h \eta_i' + h' \eta_i + \frac{i \omega z^2 h B_{it}' - 16h' \eta_i}{16 - \omega^2 z^2 h}\right) = 0
\end{equation}

Observe that equation \ref{eta-vector} contains no information for $z > z_c$, since $h(z) = 1$ and $h'(z) = 0$ in that region. Tracing this back to Eq \ref{1-form-eom}, we conclude that $\eta_1$ simply does not exist for $z > z_c$. In this interpretation, the field $B_2$ starts out life in the deep interior of the bulk and evolves continuously through the $D$-brane cap until it reaches the UV boundary. However, the field $\eta_1$ does not exist on the IR side of the $D$-brane cap - it begins its life at $z = z_c$ and evolves to the UV boundary. As $\eta_1$ started its life as the DBI worldvolume gauge field (which was then mixed together with other bulk fields to obtain the physical spectrum), it makes sense that it only exists where the $D$-brane is present. 

\begin{figure}[!ht]
	\includegraphics[width=20cm,
	height=20cm,
	keepaspectratio]{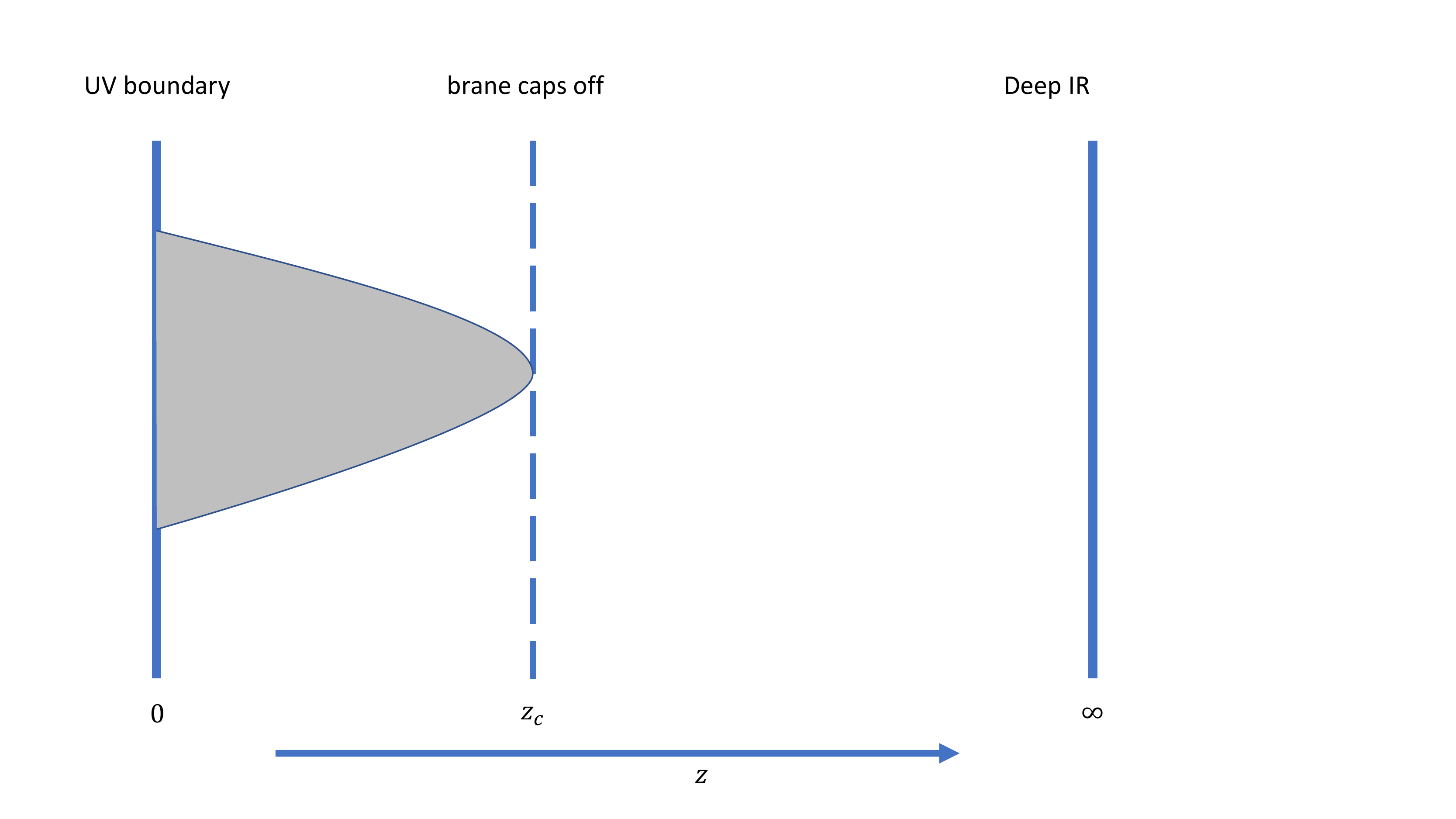}
	\caption{Brane caps off at $z = z_c$}\label{brane cap}
\end{figure}

\begin{figure}[!ht]
	\includegraphics[width=20cm,
	height=20cm,
	keepaspectratio]{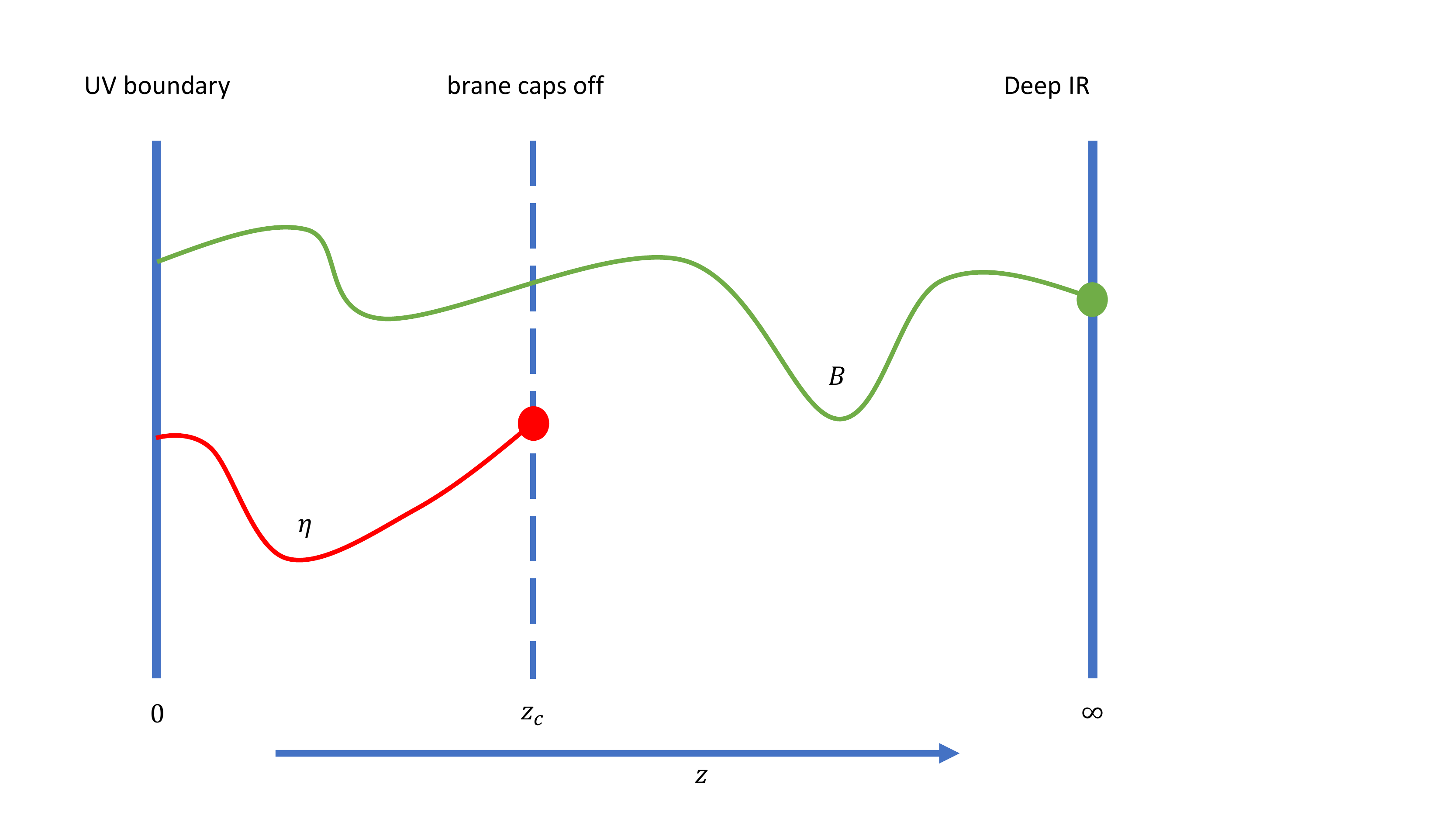}
	\caption{$B_2$ (in green) evolves continuously from the IR to the UV boundary; $\eta_1$ (in red) is ``born'' at the brane cap and evolves to the UV}
\end{figure}

However, we now need to understand how to evolve the existing fields through the transition at $z = z_c$. Imposing continuity of the $\eta_1$ equation of motion yields a useful boundary condition. As $z \to z_c$ from below we have $h(z) \to 1$ and $h'(z) \to 2\mu/z_c$, so we obtain
\begin{equation*}
	dz \wedge \star \left(-d\eta_1 + B_2 + T_2 + \frac{2 \mu}{z_c} \eta_1 \wedge dz\right) = 0
\end{equation*}
i.e.
\begin{equation}
	B_{\mu z} + T_{\mu z} - \partial_\mu \eta_z + \left(\partial_z + \frac{2\mu}{z_c}\right) \eta_\mu = 0
\end{equation}
With our gauge choice, the relevant boundary condition at the cap is given by 
\begin{equation}
	\eta_i' = - \left(\frac{i \omega B_{it}' - 2 \omega^2 \mu \eta_i}{16 - \omega^2}\right); \quad z = z_c \ . 
\end{equation}
Finally, if we expand the dynamical equations of motion in the UV, from the asymptotic behavior of the fields we can read off the dual conformal dimensions (using e.g. \cite{CasalderreySolana:2011us})
\begin{subequations}
	\begin{align}
		\Delta_{\eta} &= 3		\\
		\Delta_{b}	&= 2 + 4 \sqrt{1+\mu}
	\end{align}
\end{subequations}
As $\eta$ is dual to the conserved baryon current, its dimension is fixed at $3$ as expected; $B$ is dual to a massive tensor mode that does not have a simple universal interpretation. 

To solve the equations of motion, we now Wick-rotate to Euclidean signature by setting $\omega = i \tilde{\omega}$. The equations of motion become
\begin{subequations}
	\begin{alignat}{2}
		&	(1-h) \left(\eta_i'' - \frac{1}{z} \eta_i' - \tilde{\omega}^2 \eta_i \right) + h^{-1}h' \left(-h \eta_i' - h' \eta_i + \frac{\tilde{\omega} z^2 h B_{it}' + 16h' \eta_i}{16 + \tilde{\omega}^2 z^2 h}\right) && = 0	\\
		&	zh \left[\frac{z(B_{it}' - \tilde{\omega} h' \eta_i)}{16 + \tilde{\omega}^2 z^2 h} \right]' - B_{it} && = 0
	\end{alignat}
\end{subequations}
We can also rewrite the equations in terms of a dimensionless holographic radial coordinate and frequency by defining
\begin{subequations}
	\begin{align}
		\zeta	& = z/z_c					\label{zeta-def}	\\	
		w		& = \tilde{\omega} \;  z_c	\label{w-def}
	\end{align}
\end{subequations}
Dropping the subscripts $i,t$ and exploiting the fact that $z \partial_z = \zeta \partial_\zeta$, we have
\begin{subequations}
	\begin{alignat}{2}
		&	(1-h) \left(\frac{d^2 \eta}{d\zeta^2} - \frac{1}{\zeta}\frac{d\eta}{d\zeta} - w^2 \eta \right)
		+ h^{-1} \frac{dh}{d\zeta} \left(-h \frac{d\eta}{d\zeta} - \frac{dh}{d\zeta} \eta
		+  \frac{z_c w \zeta^2 h \frac{dB}{d\zeta} + 16  \frac{dh}{d\zeta} \eta}{16 +  w^2 \zeta^2 h}\right) && = 0	\\
		&	\zeta h \frac{d}{d \zeta} \left[\frac{\zeta }{16 + w^2 \zeta^2 h} \left(z_c \frac{dB}{d\zeta} - w \frac{dh}{d\zeta} \eta\right) \right] - z_c B && = 0
	\end{alignat}
\end{subequations}
Note that instances of $z_c$ remain - this is to be expected since it is precisely the mass scale $m_{\text{meson}} = z_c^{-1}$ which breaks conformal invariance of the dual field theory. However, the factors of $z_c$ appear only when multiplied by $B$. Hence we can define $b = z_c B$, so that

\begin{subequations}
	\begin{alignat}{2}
		&	(1-h) \left(\frac{d^2 \eta}{d\zeta^2} - \frac{1}{\zeta}\frac{d\eta}{d\zeta} - w^2 \eta \right)
		+ h^{-1} \frac{dh}{d\zeta} \left(-h \frac{d\eta}{d\zeta} - \frac{dh}{d\zeta} \eta
		+  \frac{w \zeta^2 h \frac{db}{d\zeta} + 16  \frac{dh}{d\zeta} \eta}{16 +  w^2 \zeta^2 h}\right) && = 0	\\
		&	\zeta h \frac{d}{d \zeta} \left[\frac{\zeta }{16 + w^2 \zeta^2 h} \left(\frac{db}{d\zeta} - w \frac{dh}{d\zeta} \eta\right) \right] - b && = 0
	\end{alignat}
\end{subequations}

The boundary condition at the cap is given in Euclidean signature by
\begin{equation}
	\frac{d \eta}{d \zeta} = \frac{w \frac{db}{d\zeta} - 2 \mu w^2 \eta}{16 + w^2}
\end{equation}

\subsubsection{2-form}
The above set of equations is a closed system that can be conveniently numerically solved. However we are ultimately interested in studying the behavior of the system in the $U(N)$ duality frame, in which the physics is encoded in the fields $\sP_2$ and $\sA_2$ rather than $\eta_1$ and $\tau_1$. To relate them, we note that in the UV ($z \to 0, \; dh = 0$), we can match the fields using \eqref{Q3def} to get $d\mathcal{P}_2 = \frac{\kappa^2 \mu}{1+\mu} \star d\eta_1$. After a Wick rotation we can fix some UV scale $z_\Lambda$ to get
\begin{subequations}
	\begin{align}
		\frac{w}{z_c} \; \mathcal{P}_{12}(z_\Lambda) &= \alpha \; \frac{\eta_3'(z_\Lambda)}{z_\Lambda}	\\
		z \; \mathcal{P}_{12}'(z_\Lambda) & =  \alpha \; \frac{w}{z_c} \; \eta_3(z_\Lambda)
	\end{align}
\end{subequations}
where $\alpha = \frac{16 \mu}{(1+\mu)R}$.
As in \cite{Hofman:2017vwr}, in the UV we have
\begin{equation}
	\mathcal{P}_{jk} \sim p_{jk} + J_{jk} \log z, \quad z \to 0 \label{formofLog} 
\end{equation}
We may also directly verify that the leading order asymptotic behaviour of $\eta_3(z)$ is given by
\begin{equation}
	\eta_3(z) \sim \eta_0 + \eta_2 z^2 + \bar{\eta}_2 z^2 \log z, \quad z \to 0
\end{equation}
Hence matching these components at the cutoff we find that 
\begin{subequations}
	\begin{align}
		J_{12}											& = \alpha \; \frac{w}{z_c} \; \eta_0		\\
		\frac{w}{z_c} (p_{12} + J_{12} \log z_\Lambda)	& = \alpha \left(2 \eta_2 + \bar{\eta}_2 + 2 \bar{\eta}_2 \log z_\Lambda\right)
	\end{align}
\end{subequations}
Consistency of the unambiguous coefficients of $\log z_\Lambda$ fixes the coefficient $\bar{\eta}_2$ to be
\begin{equation}
	\bar{\eta}_2 = \frac{1}{2} \left(\frac{w}{z_c}\right)^2 \eta_0
\end{equation}
We now turn to the interpretation of the logarithm in Eq \eqref{formofLog}. As explained in detail in \cite{Hofman:2017vwr}, this logarithm arises from the fact that the double-trace coupling $J^2$ is marginally irrelevant. This marginal irrelevance breaks conformality, and more information must be given to specify the theory. (Indeed, the only conformal theory with a continuous 1-form symmetry in four dimensions is free Maxwell electrodynamics \cite{Hofman:2018lfz,Cordova:2018cvg}). This information can be given in the form of the value of the double-trace coupling $\frac{1}{\theta} J^2$ at a particular scale. (Note that in this strongly coupled model one can now identify $\theta$ with the gauge coupling of the $U(1)$ sector $g_1$ in \eqref{deformLag2}). 

Following the algorithm in \cite{Hofman:2017vwr}, we can now determine the source $p_{12}$ by
\begin{equation}
	p_{12}= \mathcal{P}_{12}(z_\Lambda) - \frac{J_{12}}{\theta} = 2\alpha \; \frac{z_c}{w} \; \eta_2 + J_{12} \log \bar{z}_*
\end{equation}
Here the scale $\bar{z}_*$ is given by
\begin{equation}
	\bar{z}_* \equiv e^{1/2} \; z_* \equiv e^{1/2} \; z_\Lambda \; e^{-1/\theta}
\end{equation}
The value of this scale should be understood as the Landau pole where the theory breaks down; as $\th > 0$, we note that it is an extremely small scale, much smaller than the cutoff. 
Concretely, we can numerically extract the 2-point function content $f_{JJ}(\omega)$ by solving the equations of motion for $B_{3t}$ and $\eta_{3}$. See Appendix \ref{code-appendix} for further details of this method.

\subsubsection{Results}
Here we present a plot of the numerically calculated Green's function as a function of $w = \omega z_c$ for various values of the dimensionless number $\gamma \equiv z_c/\bar{z}_*$, i.e. the meson mass in units of the Landau pole scale. 

Note that at weak coupling the mass gap is given by the bare flavor mass $m_F$. However at strong coupling the mass gap is the mass of the lightest meson which is given by $\frac{1}{z_c} = \frac{1}{2\pi}\frac{\sqrt{\lambda}}{m_F}$, where here $m_F$ should be understood as the coefficient of the relevant mass deformation in the UV. We have thus chosen to plot the result in units of the physical meson mass $z_c$. 
\begin{figure}[!ht]
	\includesvg[width = \textwidth]{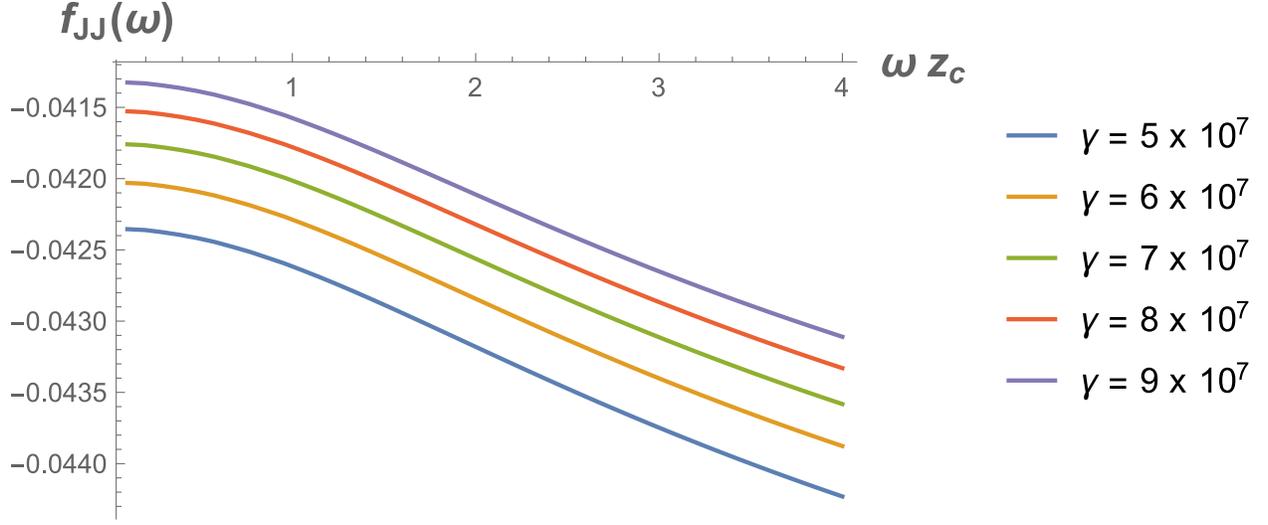}
	\caption{The symmetry current correlator at strong coupling as a function of $w = \omega z_c$ for various masses, computed numerically using holography. For this plot we set $\bar{z}_* = 10^{-8}$.}
	\label{fig:numerical} 
\end{figure}
We observe that the asymptotic behaviour is as expected: for small $w$ the leading order contribution is a constant which depends on $\gamma$. For large $w$ we expect logarithmic behaviour, but this is difficult to see explicitly because we cannot numerically access the solution for an exponential range of values of $w$.

\subsection{Comparison to weak coupling}
Here we will try to compare the functional form of the results above to a weak-coupling computation. By weak-coupling, we mean that we will take the non-Abelian t'Hooft coupling $\lambda$ to zero; however we will keep fixed the Landau pole associated with the $U(1)$ factor. Note that in the $\lambda$ to zero limit, the $U(1)$ sector of the field theory is effectively super QED with $N_f$ flavours, i.e. a $U(1)$ gauge field $a_1$ coupled to $N_f$ Dirac fermions and $N_f$ complex scalars of mass $m$ with coupling constant $g_1$. Up to a normalisation, the current associated with the $1$-form global symmetry is $J_b = \star f$.

The current-current correlator can be shown to be
\begin{equation}
	\braket{\tilde{J}_b^{\mu \nu}(k)\tilde{J}_b^{\rho \sigma}(-k)}
	= \epsilon^{\mu \nu \alpha \beta} \epsilon^{\rho \sigma \gamma \delta} k_\alpha k_\gamma \; \tilde{\Delta}_{\beta \delta}(k)
\end{equation}
where $\tilde{\Delta}_{\mu \nu}(k)$ is the loop-corrected photon propagator in momentum space. We are interested in the purely spatial components $\braket{\tilde{J}_b^{xy}(k)\tilde{J}_b^{xy}(-k)}$.

\begin{figure}[!ht]
	\includegraphics[width=5cm,
	height=10cm,
	keepaspectratio]{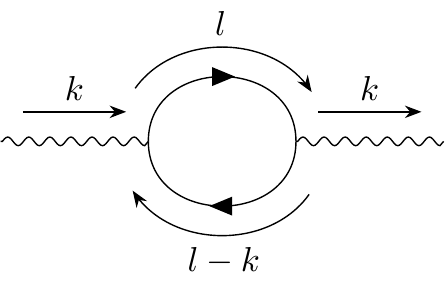}
	\caption{Fermion loop diagram contributing to correction of photon propagator}
	\label{fermion-loop-photon}
\end{figure}

The contributions to the photon propagator $\Delta_{\mu \nu}$ arise from resumming scalar and fermion loops as in Figure \ref{scalar-loop-photon} and Figure \ref{fermion-loop-photon}:

\begin{figure}[!ht]
	\includegraphics[width=5cm,
	height=10cm,
	keepaspectratio]{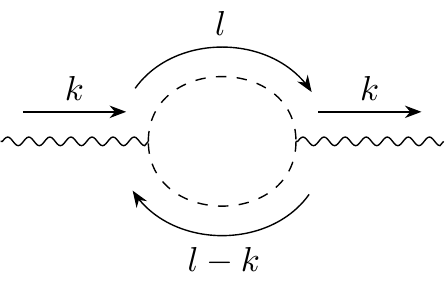}
	\caption{Scalar loop diagram contributing to correction of photon propagator}
	\label{scalar-loop-photon}
\end{figure}

This is a textbook calculation - see e.g. \cite{srednicki2007quantum} for a reference which matches our conventions. We use dimensional regularisation in the $\overline{MS}$ renormalisation scheme and put momentum purely in the time direction. This allows us to write
\begin{equation}
    \braket{\tilde{J}_b^{xy}(\omega)\tilde{J}_b^{xy}(-\omega)}
    =\left\{
    1-\frac{N_f \; g_1^2 (\mu)}{4\pi^2}\int_0^{1/2} dy\; (1-2y^2)\log\left[\frac{1 + (1/4-y^2) \; \hat{\omega}^2}{(\mu/m)^2}\right] + \mathcal{O}(g_1^4)
    \right\}^{-1}
\end{equation}
where $\mu$ is an arbitrary mass scale and we define a dimensionless number by $\hat{\omega} \equiv \omega/m$. The coupling $g_1$ runs logarithmically with the energy scale. Let's fix the coupling $g_1$ at some UV scale $\mu_\Lambda$ to be $g_R$. Then the Landau pole scale $\mu^*$ at which the renormalized coupling $g_1$ diverges is related to $\mu_\Lambda$ by
\begin{equation}
    \mu^* = \mu_\Lambda \; e^{1/\chi}
\end{equation}
where here $\chi$ is given by 
\begin{equation}
    \chi = \frac{5 N_f \; g_R^2}{24\pi^2}
\end{equation}
Note that we have combined the fermion contribution of $N_f \; g_R^2 / (6 \pi^2)$ with the scalar contribution of $N_f \; g_R^2 / (24 \pi^2)$. Here $\mu^{*}$ is the physical scale which we should identify with the holographic Landau pole $z^*$ when comparing the two theories.

This gives an expression for the current-current correlator in terms of the Landau pole scale and the double-trace coupling as 

\begin{equation}
	\braket{\tilde{J}_b^{xy}(\omega)\tilde{J}_b^{xy}(-\omega)}^{-1}
	= \frac{6}{5} \chi \int_{-1/2}^{1/2} dy \; (1-2y^2) \; \frac{1}{2} \log \left[\frac{1 + (1/4-y^2)\; \hat{\omega}^2}{(\mu^*/m)^2}\right]
\end{equation}

\begin{figure}[!ht]
	\includesvg[width = 15cm]{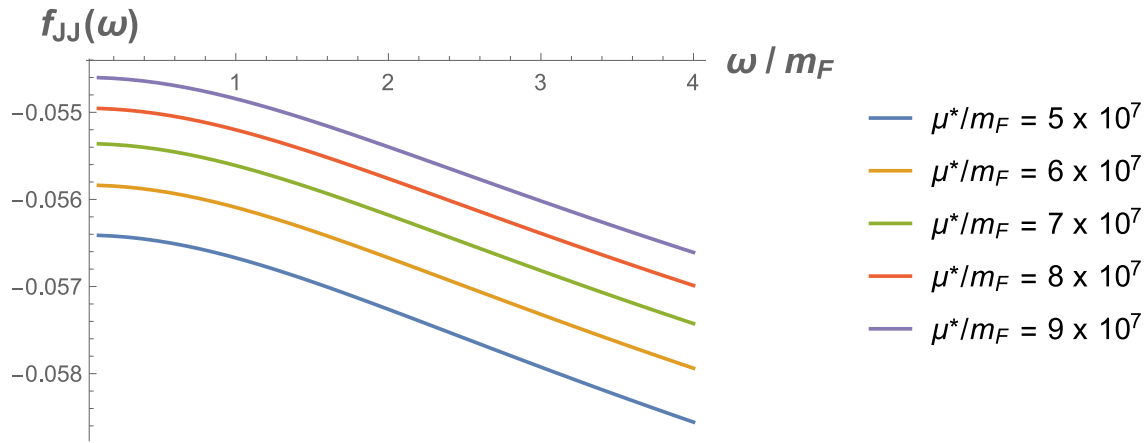}
	\caption{The symmetry current correlator at weak coupling as a function of $\hat{\omega} = \omega / m_F$, computed analytically in perturbation theory}
	\label{Greens-analytic}
\end{figure}

See Figure \ref{Greens-analytic} for a plot of the correlator at weak coupling. As we can see, the weak-coupling and strong-coupling plots are extremely similar: they approach a constant for small $\omega$ in relation to the relevant mass scale (as dictated by the spontaneous symmetry breaking), and diverge logarithmically for large $\omega$ (as dictated by the running of the coupling at large frequencies). It is curious to note that these two properties appear to be sufficient to control the correlator at all scales, giving the same dynamical behavior from strongly coupled gravity and from weakly coupled Feynman diagrams. 

\vspace{0.2in}
{\bf Conclusion: } 
In this work, we have studied the realisation of 1-form symmetries in perhaps the simplest holographic model in which such a symmetry could be spontaneously broken; along the way we have clarified some aspects of the interplay between $0$-form baryon number symmetry and the $1$-form $\mathbb{Z}_{N}$ symmetry in $SU(N)$ gauge theory. We identified the charged line operator and verified the expected behavior of the current-current correlation function, demonstrating the existence of the expected Goldstone mode. We can identify various directions for future research. It would be very interesting to extend this study to finite temperatures, where we could expect to make contact with recent symmetry-based formulations of magnetohydrodynamics \cite{Grozdanov:2016tdf}. In a more formal direction, it would also be very interesting to understand the bulk holographic dual of the colour-flavour-center symmetry identified in \cite{Cherman:2017tey}.

\vspace{0.2in}   \centerline{\bf{Acknowledgements}} \vspace{0.2in} We thank N. Poovuttikul for discussions. NI and KM are supported in part by the STFC under consolidated grant ST/L000407/1.  

\clearpage
\begin{appendix}

\section{Conventions and differential form identities}
\label{conventions-appendix}
In this work we normally use $M,N$ to refer to 5d bulk indices, $\mu,\nu$ to refer to 4d field theory bulk indices, and $i,j$ to refer to 3d spatial indices. $\alpha, \beta$ refer to $D7$-brane worldvolume coordinates and $A, B$ refer to 10d target space indices.

Our our conventions for differential forms are those of \cite{carroll2004spacetime}, and we record some useful identities below:
\begin{align}
d(\om_p \wedge \eta_q) & = d\om_p \wedge \eta_q + (-1)^p \om_p \wedge d\eta_q \\
\om_p \wedge \eta_q & = (-1)^{pq} \eta_q \wedge \om_p \\
\om_p \wedge \star\eta_p & = \eta_p \wedge \star\om_p
\end{align}

The square of the Hodge star acting on a $p$ form in $n$ dimensions on a metric with $s$ minus signs in its eigenvalues is
\begin{equation}
\star^2 = (-1)^{s+p(n-p)} \ . 
\end{equation}

In particular, in Lorentzian signature in 4d acting on a 2-form, we have $\star_4^2 = -1$.

To translate between our expressions involving forms and our expressions involving components, we can use the identity
\begin{equation}
	A_p^2 = |A_p|^2 \; \epsilon
\end{equation}
where $\epsilon$ is the volume form associated with the metric determinant $g$
\begin{equation}
	\epsilon = \star 1 = \sqrt{|g|} \; d^n x
\end{equation}
As in \cite{Polchinski:1998rr}, we define
\begin{equation}
	|A_p|^2 = \frac{1}{p!}A_{\mu_1 \ldots \mu_p}A^{\mu_1 \ldots \mu_p}
\end{equation}
and we use the shorthand
\begin{equation}
	A_p^2 \equiv A_p \wedge \star A_p
\end{equation}
The integral of an $n$-form $\Omega$ over an $n$-dimensional manifold $\mathcal{M}$ of signature $s$ is defined by 
\begin{equation}
	\int_\mathcal{M} \Omega \equiv \int_{\mathbb{R}^n} (-1)^s (\star \Omega) \; \epsilon = \int_{\mathbb{R}^n} d^n x \; \sqrt{|g|}\; (-1)^s (\star \Omega)
\end{equation}
So in particular, for a $p$-form we have
\begin{equation}
	\int_\mathcal{M} A_p^2 = \int_{\mathbb{R}^n} d^n x \; \sqrt{|g|}\; |A_p|^2
\end{equation}


\section{Normalisations}
To translate between field theory quantities and bulk quantities we use the holographic dictionary \cite{Polchinski:2010hw}
\begin{equation}
\frac{R^4}{l_s^4} = \lambda \equiv g_{YM}^2 N_c = 4\pi g_s N_c
\end{equation}
\subsection{Kinetic Terms}
The kinetic terms for $B_2$ and $C_2$ are of the form
\begin{equation}
S_{\text{kin}} = -\int_{\text{AdS}_5}{\left(\frac{1}{2}\mathcal{N}_B^2 \; H_3^2 + \frac{1}{2}\mathcal{N}_C^2 \; G_3^2 \right)}
\end{equation}
and our task is to find the factors $\mathcal{N}_B$ and $\mathcal{N}_C$.

Consider the type IIB low energy supergravity action in the NS-NS sector\footnote{To match the conventions of \cite{Polchinski:1998rr} with ours we have $\int{d^{10}x \sqrt{-G} \; |F_{p+1}|^2} =	\int{F_{p+1}^2}$.} (see e.g. \cite{Polchinski:1998rr}):
\begin{equation}
	S_{\text{NS}} = \frac{1}{2 \kappa_{10}^2} \int{d^{10}x \sqrt{-G} \; e^{-2 \Phi} \left(R + 4 \partial_\mu \Phi \partial^\mu \Phi - \frac{1}{2} |H_3|^2 \right)}
\end{equation}
where $G$ is the $10$-dimensional metric in the string frame, $\Phi$ is the dilaton field, $R$ is the Ricci scalar and $\kappa_{10}$ is the gravitational coupling in $10$ spacetime dimensions given by $2\kappa_{10}^2= (2\pi)^7 l_s^8$.

If we choose the dilaton to be constant with $e^\Phi = g_s$ and choose the string frame metric to be the usual metric on $\text{AdS}_5 \times S^5$, the relevant term is
\begin{equation}
	S_{\text{NS}}	= \frac{1}{(2\pi)^7 l_s^8 g_s^2} \int_{\text{AdS}_5 \times S^5}{\left(- \frac{1}{2} H_3^2 \right)}
\end{equation}
We dimensionally reduce on the $S^5$ which yields a factor of $V_5 = \pi^3 R^5$
\begin{equation}
	S_{\text{eff}} = -\frac{R^5}{128 \pi^4 l_s^8 g_s^2} \int_{\text{AdS}_5}{\left(\frac{1}{2} H_3^2 \right)}
\end{equation}
where now the integral is taken only over the $\text{AdS}_5$ directions. We thus conclude that
\begin{equation}
\mathcal{N}_B^2 = \frac{R^5}{128 \pi^4 g_s^2 l_s^8} = \frac{N_c^2}{8 \pi^2 R^3}
\end{equation}
The analysis for the R-R kinetc term is similar. The supergravity action in the R sector is
\begin{equation}
S_{\text{R}} = - \frac{1}{4 \kappa_{10}^2} \int{d^{10}x \sqrt{-G}\left(|F_1|^2 + |\hat{G}_3|^2 + \frac{1}{2}|\tilde{F}_5|^2\right)}
\end{equation}
where the relevant quantity for us is $\hat{G}_3 \equiv G_3 - C_0 \wedge H_3$ and $G_3 = dC_2$ is the R-R field strength. Setting $C_0 = 0$ we have the term
\begin{equation}
S_{\text{R}} = -\frac{R^5}{128 \pi^4 l_s^8} \int_{AdS_5}{\left(\frac{1}{2}G_3^2\right)}
\end{equation}
after compactifying on the $S^5$. By comparing with the original action we can identify
\begin{equation}
\mathcal{N}_C^2 = \frac{R^5}{128 \pi^4 l_s^8} = \frac{\lambda^2}{128 \pi^4 R^3}
\end{equation}
as promised.
\subsection{Chern-Simons Term} \label{app:CS} 
\label{Chern-Simons appendix}
Suppose we have a Chern-Simons term in the action of the form
\begin{equation}
	S_{\text{CS}} = \frac{k}{2\pi}\int{B_2 \wedge G_3} = \kappa \; \mathcal{N}_B \; \mathcal{N}_C \int{B_2 \wedge G_3}
\end{equation}
The coupling to $D1$-branes and $F1$-strings respectively is $S_{\text{D1}} = \mu_1 \int{C_2}$ and $S_{\text{F1}} = \frac{1}{2\pi l_s^2}\int{B_2}$, where $\mu_1^{-1} = 2\pi l_s^2$ is the tension of a $D1$-brane.
A higher-form Dirac quantisation condition gives
\begin{equation}
	\frac{\mu_{1}}{2 \pi} \int_{S_{3}} F_{3} \in \mathbb{Z} \label{D5brane} 
\end{equation}
For a magnetic monopole of unit charge we have (in $d = 10$)
\begin{equation}
dG_3 + \frac{2\pi}{\mu_1} \delta_4(W) = 0
\end{equation}
where $W = S^5 \times L$ is the 6d worldvolume of the $D5$-brane sourcing the monopole and $L$ is the worldline of the monopole, i.e.a timelike curve in $\text{AdS}_5$.

By taking the wedge with $d\Omega_5 \wedge \Xi_1$ and integrating over all $10$ dimensions we obtain
\begin{equation}
\int_{\text{AdS}}{dG_3 \wedge \Xi_1} + \frac{2\pi}{\mu_1} \int_L \Xi_1 = 0
\end{equation}
which allows us to write the gauge variation of the Chern-Simons term as
\begin{equation}
\delta S_{\text{CS}} = \frac{k}{\mu_1} \int_L{\Xi_1}
\end{equation}
This must be cancelled by the gauge variation of $M$ $F_1$ strings which end on the worldline $L$,
\begin{equation}
M \delta S_{F1} = \frac{M}{2\pi l_s^2} \int_{F1}{\delta B_2} =  \frac{M}{2\pi l_s^2} \int_{L}{\Xi_1}
\end{equation}
We identify the integer $M$ with the number of colours in the field theory $N_c$, as in \cite{Witten:1998xy}.

Hence 
\begin{equation}
k = \frac{\mu_1 N_c}{2 \pi l_s^2} = \frac{N_c}{4\pi^2 l_s^4} = \frac{N_c \lambda}{4\pi^2 R^4}
\end{equation}
This gives
\begin{equation}
\kappa = \frac{k}{2\pi \mathcal{N}_B \mathcal{N}_C} = \frac{4}{R}
\end{equation}
\subsection{DBI term}
\label{DBI appendix}
Here we will describe the basic setup to add a $D7$-brane to $\text{AdS}_5 \times S^5$ by wrapping an $S^3$ around the $S^5$. We will follow a similar approach to \cite{Karch:2007pd}. The final result will be a contribution to the action of
\begin{equation}
	-\mathcal{N}_B^2 \int{\left[\frac{1}{2}\kappa^2 \mu f(z) \left(B_2 + \frac{2 \pi R^2}{\sqrt{\lambda}}F_2\right)^2\right]}
\end{equation}
where the factor $\mu$ and the function $f(z)$ will be determined.

The  $10d$ string frame metric $G_{AB}$ is given by
\begin{equation}
ds^2 = G_{AB}dX^A dX^B = \frac{R^2}{z^2} (-dt^2 + dz^2 + dx^i dx^j \delta_{ij}) + R^2 d\Omega_5^2
\end{equation}
Here $i, j \in \{1,2,3\}$ are spatial indices and $A, B$ index the coordinates $z, t, x^i$ and all the angles of $S^5$. $R$ is the AdS radius and we parametrise the $5$-sphere as
\begin{equation}
d\Omega_5^2 = d\theta^2 + \cos^2{\theta} d\psi^2 + \sin^2{\theta} d\Omega_3^2,
\end{equation}
where $d\Omega_3^2$ is the standard metric for a $3$-sphere, the angle $\psi \in [0, 2\pi]$ is azimuthal and the angle $\theta$ takes values in $[0, \frac{\pi}{2}]$. This coordinate choice is analogous to the so-called Hopf coordinates on $S^3$. For our purposes, these coordinates provide a simpler way to embed a $3$-sphere inside a $5$-sphere than the usual hyperspherical coordinates.

We can embed a probe $D7$-brane into the target space by means of the DBI action:
\begin{equation}
S_{\text{DBI}} = -\tau_7 \int{d^8\xi \sqrt{-\det\left(g_{\alpha \beta} + B_{\alpha \beta} + 2\pi l_s^2 F_{\alpha \beta}\right)}}
\end{equation}
where $\xi^\alpha$ are the brane worldvolume coordinates, $g_{\alpha \beta}$ is the induced worldvolume metric on the $D7$-brane, $B_{\alpha \beta}$ are the components of the NS-NS $2$-form and $F_2 = dA_1$ is the Maxwell field strength living on the brane.

$\tau_p$ is the effective $Dp$-brane tension after absorbing the effect of the dilaton $e^\Phi = g_s$ and is given by equation (13.3.23) of \cite{Polchinski:1998rr} as
\begin{equation}
	\tau_7 = \frac{1}{g_s}\frac{1}{(2\pi)^7 l_s^8}
\end{equation}
The desired brane configurations fills all of AdS and wraps a $3$-sphere around the $S^5$, so the transverse fluctuations will be the $\theta$ and $\psi$ angles. Since $G_{AB}$ is independent of $\psi$, i.e.$\partial_\psi$ is a Killing vector of the target space metric, we will take $\psi$ to be a constant. Crucially, the $\theta$ angle will be a function of the AdS radial coordinate: $\theta = \theta(z)$.

The worldvolume metric is the pull-back of the target space metric onto the worldvolume	
\begin{equation}
g_{\alpha \beta} = \frac{\partial X^A}{\partial \xi^\alpha} \frac{\partial X^B}{\partial \xi^\beta} G_{AB}
\end{equation}
If we choose static gauge $\xi^\alpha = X^\alpha$ then we simply have	
\begin{subequations}
	\begin{align}
	g_{zz} &= G_{zz} + \left(\frac{d \theta}{dz}\right)^2 G_{\theta \theta} = G_{zz}(1 + (z\theta')^2)	\\
	g_{\alpha\beta} &= G_{\alpha\beta}, \hspace{200pt} \alpha \neq z			
	\end{align}
\end{subequations}	
Let's write $B_{\alpha \beta} + 2\pi l_s^2 F_{\alpha \beta} \equiv \tilde{B}_{\alpha \beta}$. We can expand the determinant in the DBI Lagrangian as	
\begin{align*}
\det(g_{\alpha \beta} + \tilde{B}_{\alpha \beta})
& = \det(g_{\alpha \gamma} \delta^\gamma_\beta + g_{\alpha \gamma} g_{\beta \delta} \tilde{B}^{\gamma \delta})	\\
& = \det(g_{\alpha \gamma}) \det(\delta^\gamma_\beta + g_{\beta \delta} \tilde{B}^{\gamma \delta})	\\
& = \det(G_{MN})(1 + (z\theta')^2) (R^2 \sin^2{\theta})^3 \det(\delta^\gamma_\beta + g_{\beta \delta} \tilde{B}^{\gamma \delta})
\end{align*}
For a traceless matrix $A$ we have $\det(\mathbb{1} + A) = 1 - \frac{1}{2} \Tr(A^2) + \mathcal{O}(A^3)$.

Since $g_{\beta \delta} \tilde{B}^{\gamma \delta}$ is traceless, the leading order behaviour of $\det(\delta^\gamma_\beta + g_{\beta \delta} \tilde{B}^{\gamma \delta})$ is given by
\begin{equation}
\det(\delta^\gamma_\beta + g_{\beta \delta} \tilde{B}^{\gamma \delta})
= 1 + \frac{1}{2} \tilde{B}^{\alpha \beta}\tilde{B}_{\alpha \beta} + \cdots
\end{equation}
Putting this into the DBI action and performing the integral over the unit $3$-sphere to obtain a factor of $2\pi^2$ gives
\begin{equation}
S_{\text{DBI}}
= -\frac{1}{g_s}\frac{2 \pi^2 R^3}{(2\pi)^7 l_s^8} \int{\sqrt{-G_{MN}}\; d^5 x \left[\sqrt{1 + (z\theta')^2} \; \sin^3{\theta}\left(1 + \frac{1}{4}\tilde{B}^{\alpha \beta}\tilde{B}_{\alpha \beta}\right)\right]}
\end{equation}
to quadratic order in $\tilde{B}_{\alpha \beta}$.

Let's turn off the Kalb-Ramond field and the Maxwell field. Now we have an effective action of the form
\begin{equation}
S = \mathcal{N}\int{dz L[\theta(z), \theta'(z)]}
\end{equation}
where
\begin{equation}
	L = \frac{\sin^3{\theta}}{z^5}\sqrt{1 + (z\theta')^2}
\end{equation}
and we absorbed the integral over field theory directions into the overall normalisation factor $\mathcal{N}$.

Solving the Euler-Lagrange equation gives the single-valued on-shell angle
\begin{equation}
\theta(z) = \theta_c \equiv
\begin{cases} 
	\arccos(z/z_c)	& z\leq z_c \\
	0				& z > z_c
	\end{cases}
\end{equation}
Geometrically, this means that the $S^3$ wrapping the $S^5$ is of maximal size ($\theta=\pi/2$) at $z=0$ on the boundary, and the $D7$-brane vanishes ($\theta =0$) at the critical value $z=z_c$. For $z>z_c$ the $D7$-brane has no effect.	

It is straightforward to show that
\begin{equation}
	\sqrt{1 + (z\theta_c')^2} \; \sin^3{\theta_c} = f(z)
\end{equation}
where $f$ is the dimensionless scalar function given by
\begin{equation}
	f(z) = 
	\begin{cases} 
	1 - (z/z_c)^2	& z\leq z_c \\
	0				& z > z_c
	\end{cases}
\end{equation}
Using the holographic dictionary we find that $2\pi l_s^2 = \frac{2 \pi R^2}{\sqrt{\lambda}}$. In general to add $N_f$ flavours we simply add $N_f$ probe $D7$-branes. After holographic renormalisation this gives an overall contribution to the action of
\begin{equation}
N_f S_{\text{DBI}}
= -\frac{N_c^2}{8 \pi^2 R^3} \int{\left[\frac{1}{2}\left(\frac{4}{R}\right)^2 \frac{N_f}{N_c} \frac{\lambda}{32 \pi^2} f(z)\left(B_2 + \frac{2 \pi R^2}{\sqrt{\lambda}}F_2\right)^2\right]}
\end{equation}
from which we can read off that
\begin{equation}
	\mu = \frac{N_f}{N_c} \frac{\lambda}{32 \pi^2}
\end{equation}
as expected.

\subsection{Couplings of other branes}
Here we work out the couplings to various other bulk objects in our normalisation. We will sometimes make use of the form delta function $\delta_{\sM_p}(x)$. This is a delta function that is nonzero only if $x$ is on the submanifold $\sM_p$; more precisely it is a $d-p$-form such that $\delta_{\sM_p}(x) = 0$ if $x \notin \sM_p$ and the integral over any $p$-form $C_p$ satisfies:
\begin{equation}
\int_{\mathbb{R}^d} \delta_{\sM_p} \wedge C_p = \int_{\sM_p} C_p \ . 
\end{equation}
\subsubsection{Wrapped \texorpdfstring{$D5$}{D5}-brane: baryon vertex} 
We recall the action:
\begin{equation}
	S = - \frac{N_c^2}{8 \pi^2 R^3}\int \left[\frac{1}{2} H_3^2 + \frac{1}{2} \kappa^2 (B_2 + T_2)^2 \right]
\end{equation}
where $T_2 = d\tau_1$. Recall from \eqref{tildeCdef} also that $\tau_1$ is related to the original RR and NS 2-forms as
\begin{equation}
 d \tau_1 = \frac{\lam}{4\pi N_c \ka} \star dC_2 - B_2 \label{dualityrel_app} 
 \end{equation} 
We now start with a $D5$-brane wrapped on the $S^5$, and would like to determine its coupling to the field $\tau_1$; in other words we add to the action a term
\begin{equation}
q_{\tau} \int_{L} \tau_1
\end{equation}
and would like to determine the coefficient $q_{\tau}$. For simplicity, study a configuration with $B_2 = 0$; varying the action with respect to $\tau_1$ we have
\begin{equation}
\frac{N_c^2 \ka^2}{8 \pi^2 R^3} d \star d \tau_1 = q_{\tau} \delta_{L}(x) 
\end{equation}
Integrating both sides of this over a ball with boundary $S^3$ that intersects the worldline $L$, we have
\begin{equation}
\frac{N_c^2 \ka^2}{8 \pi^2 R^3} \int_{S^3} \star d\tau_1 = q_{\tau}
\end{equation}
Now use $G_3 = dC_2$ and insert \eqref{dualityrel_app} to find that for a minimally charged $D5$-brane as in \eqref{D5brane} we have
\begin{equation}
q_{\tau} = \frac{N_c}{2\pi l_{s}^2}
\end{equation}
This is $N_c$ times the ``unit charge'' of a single $F$-string in the appropriate units, as we discuss in the bulk of the text. 

\subsubsection{ Wrapped \texorpdfstring{$D5$}{D5}-brane with boundary: DBI monopole}
We now discuss a different bulk object, though also arising from a wrapped $D5$-brane, here wrapping a half $S^4$ and ending on the D7 flavour branes.  The geometry is described around \eqref{S5repeat}. Here we work out the precise charges; the computation outlined here is a higher-dimensional analogue of the calculations in \cite{Iqbal:2014cga}. The relevant parts of the bulk action are

\begin{equation}
N_{F} T_{7} \int_{D7} 2\pi l_s^2 \; F_{2} \wedge C_{6} + T_{5} \int_{D5} C_{6} + \cdots
\end{equation}
We study the case with $N_f = 1$. We study the configuration where the $D5$-brane has a boundary $\p D5$ ending on the $D7$-brane. This boundary means that the coupling to $C_{6}$ alone is no longer gauge invariant; indeed if we now do a 5-form gauge transformation of the RR 6-form $C_6$,  $C_{6} \to C_{6} + d\Lambda_{5}$, we find that gauge-invariance requires that
\begin{equation}
T_{7} \; 2\pi l_{s}^2 \; d F_{2} = -T_{5} \; \delta_{\p D5}(x)
\end{equation}  
and thus that if we consider an $S^2$ that surrounds $\p D5$ on the $D7$-brane worldvolume, we have that
\begin{equation}
\int_{S^2} F_{2} = 2\pi 
\end{equation}
where we have used that $\frac{T_5}{T_{7} (2\pi l_s^2)} = 2\pi$. Thus the edge of the wrapped brane couples magnetically to the DBI worldvolume field. As expected, this is the magnetic flux that saturates the Dirac quantisation condition, where the conjugate electric charge is viewed as the endpoint of an F-string ending on the $D7$-brane. 

\subsubsection{\texorpdfstring{$D1$}{D1} string}
Here we work out the coupling of the $D1$ string to $\tilde{C}_1$, which is the magnetic dual of the RR 2-form $C_1$.  We begin with the relevant part of the effective 5d kinetic term for $C_2$ from \eqref{bulkac}, which is
\begin{equation}
S = - \frac{\lam^2}{128 \pi^3 R^3} \int_{AdS_{5}} \ha (dC_2)^2 + \cdots
\end{equation}
From here we and the coupling to a $D1$ string used in \ref{app:CS} we find that the equation of motion in the presence of a $D1$ source is
\begin{equation}
\int_{S^2} \star d C_2 = \frac{64 \pi^3}{l_s^2}\frac{R^3}{\lam^2} 
\end{equation}
where the integral is taken over an $S^2$ that surrounds the $D1$ brane. Next, using the relation between $C_2$ and $\tilde{C}_1$ in \eqref{tildeCdef} in a configuration where $B = 0$, we find that
\begin{equation}
\int_{S^2} dC_1 = \frac{16\pi^2 l_s^2}{N_{C} R} 
\end{equation}
Restoring the factor of $\ka^{-1}$ this reduces to \eqref{D1coupling} quoted in the main text. 

\section{Numerical Solution}
\label{code-appendix}
Here we give a brief explanation of the numerical procedure used to obtain the spectral function in Figure \ref{fig:numerical}.

An overview: the equations of motion for $B_2$ and $\eta_1$ are numerically solved twice, each time with different boundary conditions. The solution for $\eta_1$ is dualised to a solution for $\mathcal{P}_2$ in the UV. The solutions for $B_2$ and $\mathcal{P}_2$ are then used to construct the Green's function $f_{JJ}$. For concreteness, we work with the components $B_{3t}$ and $\eta_3$.
\subsection{Equations of motion}
The boundary conditions used are the values of the fields $B$ and $\eta$ at the $D7$-brane cap $\zeta = 1$. This then fixes the derivatives of the fields as follows. The derivative of $B$ is determined by solving the equation of motion in the range $1 < \zeta < \infty$ and the derivative of $\eta$ is determined by imposing continuity as $\zeta \to 1$ from below.

In the region $1 < \zeta < \infty$, the equation of motion for $B$ is considerably simpler. In fact, it can be reduced to a first-order nonlinear differential equation for the new field $\Sigma$ defined by
\begin{equation}
	\Sigma(\zeta) \equiv \frac{1}{B(\zeta)} \frac{dB}{d\zeta}
\end{equation}
The appropriate asymptotic boundary condition in the IR is constrained by regularity to be 
\begin{equation}
	\Sigma(\zeta) \sim - w; \quad \zeta \to \infty 
\end{equation}
This is the consequence of the asymptotics of $B$ itself:
\begin{equation}
	B(\zeta) \sim e^{-w \zeta}\; \quad \zeta \to \infty 
\end{equation}
After solving for $\Sigma$ we can read off the value of the derivative of $B$ at the cap as
\begin{equation}
	\frac{dB}{d\zeta}(1) = \Sigma(1) B(1)
\end{equation}
The coupled equations of motion are solved up to some UV-cutoff scale (a minimum value for $z_c$), at which $\eta_1$ can be straightforwardly dualised to $\mathcal{P}_2$. $\mathcal{P}_2$ corresponds to a $1$-form global symmetry, so its asymptotic form in the UV is well-understood. 

\subsection{Asymptotic analysis}
To extract the data needed for the Green's functions, we need a careful understanding of the asymptotic falloffs of various fields. In the 1-form picture, one finds the following form for the fields:
\begin{equation}
	B(z \to 0) \sim z^{-\nu} \le(b_{0,-} + z^2 b_{2,-} + \cdots\ri) + z^{\nu} \le(b_{0,+} + z^2 b_{2,+} + \cdots\ri)
\end{equation} 
Here, by the usual rules of AdS/CFT, $b_{0,-}$ is the source and $b_{0,+}$ is the response. Similarly, we may expand the field $\eta(z)$ at infinity: we find
\begin{equation}
	\eta(z \to 0) \sim \eta_0 + \eta_{2} z^2 + \bar{\eta}_2 z^2 \log z + \cdots + z^{4-\nu} \le(\eta_{-,0} + z^2 \eta_{-,2} + \cdots\ri)
\end{equation}
Here a somewhat unfamiliar role is played by the terms in $z^{4-\nu}$; these arise from the mixing between the two bulk fields. The coefficients $\eta_{-,0}, \eta_{-,2}$ are all proportional to $b_{0,-}$ and may be explicitly calculated from the asymptotic analysis of the equations of motion.

Numerically it is more practical to fit the solutions of the equations of motion to the known asymptotic form using linear regression. This ``numerical holographic renormalisation'' allows us to pick out the coefficients we need. Crucially however, we implemented the numerics using the $\zeta$ coordinate defined by $\zeta = \frac{z}{z_c}$. We can write the above asymptotic expansions in this coordinate system as
\begin{align}
	B(\zeta \to 0)
	& \sim z_c^{-\nu} \; \zeta^{-\nu} \left( b_{0,-} + z_c^2 \; \zeta^2 \; b_{2,-} + \cdots \right) + z_c^\nu \; \zeta^{\nu} \left(b_{0,+} + z_c^2 \; \zeta^2 \; b_{2,+} + \cdots \right)				\\ \nonumber 
	& = \zeta^{-\nu} \left(\hat{b}_{0,-} + \cdots \right) + \zeta^{\nu} \left(\hat{b}_{0,+} + \cdots \right)
\end{align} 
A linear regression in the $\zeta$ coordinate system will thus fit the coefficients
\begin{subequations}
	\begin{align}
		\hat{b}_{0,-} & \equiv z_c^{-\nu} b_{0,-}		\\
		\hat{b}_{0,+} & \equiv z_c^{\nu} \; b_{0,+}
	\end{align}
\end{subequations}
This scaling can be accounted for, but will anyway cancel out at the end of our calculation.

However for $\eta$ the presence of the logarithmic term in the asymptotic expansion produces a more subtle transformation. We have (after holographic renormalisation)
\begin{align}
	\eta(\zeta \to 0)
	& \sim \eta_0 + \eta_{2} z_c^2 \zeta^2  + \bar{\eta}_2 z_c^2 \zeta^2 (\log z_c + \log \zeta) + \cdots	\\ \nonumber
	& = \eta_0 + z_c^2 \left (\eta_2 + \bar{\eta}_2 \log z_c \right) \zeta^2  + \bar{\eta}_2 \; z_c^2 \; \zeta^2 \log \zeta + \cdots	\\ \nonumber
	& = \hat{\eta}_0 + \hat{\eta}_2 \zeta^2 + \hat{\bar{\eta}}_2 \zeta^2 \log \zeta + \cdots
\end{align}
Hence for $\eta$ a naive linear regression in the $\zeta$ coordinates will fit the coefficients
\begin{subequations}
	\begin{align}
		\hat{\eta}_0 & \equiv \eta_0		\\
		\hat{\eta}_2 & \equiv z_c^2 \left (\eta_2 + \bar{\eta}_2 \log z_c \right)
	\end{align}
\end{subequations}
$\bar{\eta}_2$ is given in terms of $\eta_0$ by consistency, so we can invert these transformations to obtain $\eta_0$ and $\eta_2$, i.e.the expansion coefficients in the $z$ coordinate system. These are the physically useful constituents for computing the Green's function.

We finally map these coefficients to the source and response in the $2$-form picture via
\begin{subequations}
	\begin{align}
		p & = 2\alpha \; \frac{z_c}{w} \; \eta_2 + J_{12} \log \bar{z}_*		\\
		J & = \alpha \; \frac{w}{z_c} \; \eta_0		
	\end{align}
\end{subequations}
\subsection{Source-response method}
To construct the Green's function we refer to the source-response picture, in which the Green's function is understood as acting on the source to produce a response. Labelling the fields as $I, J$, the sources as $S_I$ and the responses as $R_I$, the components $G_{IJ}$ of the Green's function are thus given by
\begin{equation}
	G_{IJ} S_J = R_I
\end{equation}
Hence for example, 
\begin{equation}
	G_{PB} S_B + G_{PP} S_P = R_P
\end{equation}
To extract $G_{PP}$, we need to obtain two sets of the source and response data, which we label as $S_I^{(1)}$ and $S_I^{(2)}$, etc. We thus obtain a straightforward matrix equation
\begin{equation}
	\begin{pmatrix}
		S_B^{(1)} & S_P^{(1)} \\
		S_B^{(2)} & S_P^{(2)}
	\end{pmatrix}
	\begin{pmatrix}
		G_{PB}\\
		G_{PP}
	\end{pmatrix}
	=
	\begin{pmatrix}
		R_P^{(1)}\\
		R_P^{(2)}
	\end{pmatrix}
\end{equation}
which we can trivially invert to find
\begin{equation}
	G_{PP} = \frac{S_B^{(1)} R_P^{(2)} - S_B^{(2)} R_P^{(1)}}{S_B^{(1)} S_P^{(2)} - S_B^{(2)} S_P^{(1)}}	\label{GPP}
\end{equation}
In our earlier notation, we have $S_B = b_{0, -}$, $R_B = b_{0, +}$, $S_P = p$, $R_P = J$ and $f_{JJ} = G_{PP}$. Hence running the numerical algorithm twice provides all the data we need to input into equation \ref{GPP} to compute the Green's function of interest.
\section{Index of symbols}
\label{symbol-index}
For the convenience of the reader, here we present a roughly alphabetical list of the symbols in this paper, a brief description, and where it is first defined. As a rule, the subscript on a form indicates the degree of the form.
\begin{enumerate}
\item $A_1$: the usual DBI worldvolume gauge field living on the flavour brane. First appears in \eqref{bulkac}, where $F_{2} = dA_1$. 
\item $\mathcal{A}_2$: the magnetic dual of the $1$-form field $\tau_1$.
\item $B_2$: the NS-NS 2-form. First appears in \eqref{bulkac}. 
\item $C_2$: the R-R 2-form. First appears in \eqref{bulkac}. 
\item $\tilde{C}_1$: the magnetic dual of the R-R 2-form $C_2$. Defined in \eqref{tildeCdef}. 
\item $f(z)$: the function describing how the brane caps off in the bulk. Defined in \eqref{f-def}.
\item $f_{JJ}(w)$: the scalar function capturing the dependence of the symmetry current two-point function on $w$. Defined in \eqref{Jcorrdef}.
\item $F_2$: the field strength of $A_1$. First appears in \eqref{bulkac}.
\item $\mathcal{F}_3$: the field strength of $\mathcal{A}_2$. Defined in \eqref{f3def}.
\item $G_3$: the field strength of $C_2$. First appears in \eqref{bulkac}.
\item $\tilde{G}_2$: the field strength of $\tilde{C}_1$. Defined in \eqref{tildeCdef}.
\item $h(z)$: a function of $f$ first defined in \eqref{h def}
\item $H_3$: the field strength of $B_2$. First appears in \eqref{bulkac}.
\item $m_F$: the fermion mass, i.e.the mass gap at weak coupling. First appears in \eqref{fermion-mass}.
\item $m_{\text{meson}}$: the lightest meson mass, $z_c^{-1}$, i.e.the mass gap at strong coupling.
\item $N_c$: the number of colours of the gauge group.
\item $N_f$: the number of flavours of fundamental matter; the number of $D7$-branes in the bulk.
\item $\mathcal{P}_2$: the magnetic dual of $\eta_1$. Defined in \eqref{Q3def}.
\item $\mathcal{Q}_3$: the field strength of $\mathcal{P}_2$. Defined in \eqref{Q3def}.
\item $R$: the AdS radius. First appears in \eqref{Poincare-coords}.
\item $T_2$: the field strength of $\tau_1$. Defined in \eqref{T2-def}.
\item $w$: the dimensionless number $\omega z_c$. Defined in \eqref{w-def}.
\item $Y_2$: the field strength of $\eta_1$. First defined in \eqref{Y2-def}.
\item $z$: the radial AdS coordinate.
\item $z_c$: the value of $z$ where the brane caps off. First appears in \eqref{f-def}.
\item $\eta_1$: a combination of $A_1$ and $\tilde{C}_1$. Defined in \eqref{etadef}.
\item $\kappa$: Factor appearing in \eqref{bulkac} equal to $\frac{4}{R}$. Defined in \eqref{kappa-def}.
\item $\mu$: the ratio of mass contributions from $A_1$ and $\tilde{C}_1$. Defined in \eqref{mu-def}.
\item $\tau_1$: a combination of $A_1$ and $\tilde{C}_1$. Defined in \eqref{taudef}.
\item $\zeta$: the dimensionless number $z/z_c$. Defined in \eqref{zeta-def}.

\end{enumerate}

\end{appendix}
\bibliographystyle{utphys}
\bibliography{all}
\end{document}